\documentclass[11pt]{article}
\usepackage[margin=1in]{geometry}
\usepackage{amsmath,amsfonts,amssymb,amsthm}
\usepackage{graphicx}
\usepackage{enumerate}
\usepackage{bbm}
\usepackage{verbatim}
\usepackage{hyperref,color}
\usepackage[capitalize,nameinlink]{cleveref}
\usepackage[dvipsnames]{xcolor}
\hypersetup{
	colorlinks=true,
	pdfpagemode=UseNone,
    citecolor=OliveGreen,
    linkcolor=NavyBlue,
    urlcolor=Magenta,
	pdfstartview=FitW
}
\usepackage{appendix}
\crefname{appsec}{Appendix}{Appendices}
\usepackage{tikz}
\usepackage[ruled,linesnumbered]{algorithm2e}
\SetKwComment{Comment}{/* }{ */}
\usepackage[colorinlistoftodos]{todonotes}

\usepackage{authblk}

\theoremstyle{plain}
\newtheorem{thm}{Theorem}[section]
\newtheorem{prop}[thm]{Proposition}
\newtheorem{lem}[thm]{Lemma}
\newtheorem{lemma}[thm]{Lemma}

\newtheorem{cor}[thm]{Corollary}
\newtheorem{conj}[thm]{Conjecture}
\newtheorem{qn}[thm]{Question}

\theoremstyle{definition}
\newtheorem{defn}[thm]{Definition}

\newtheorem*{ass*}{Assumption}
\newtheorem*{notn}{Notation}

\theoremstyle{remark}
\newtheorem{rmk}[thm]{Remark}
\newtheorem{rem}[thm]{Remark}

\crefname{lem}{Lemma}{Lemmas}
\crefname{thm}{Theorem}{Theorems}
\crefname{defn}{Definition}{Definitions}
\crefname{fact}{Fact}{Facts}
\crefname{clm}{Claim}{Claims}
\crefname{prop}{Proposition}{Propositions}
\crefname{algocf}{Algorithm}{Algorithms}

\newcommand{\E}{\mathbb{E}}

\DeclareMathOperator{\good}{good}
\DeclareMathOperator{\bad}{bad}
\DeclareMathOperator{\set}{set}
\DeclareMathOperator{\failed}{failed}
\DeclareMathOperator{\coupled}{coupled}
\DeclareMathOperator{\agree}{agree}
\DeclareMathOperator{\unsat}{unsat}

\DeclareMathOperator{\vbl}{vbl}
\DeclareMathOperator{\Lin}{Lin}
\DeclareMathOperator{\Poi}{Poi}
\newcommand{\norm}[1]{\left\lVert #1 \right\rVert}
\newcommand{\ceil}[1]{\left\lceil #1 \right\rceil}

\newcommand{\poly}{\mathrm{poly}}
\newcommand{\dist}{\mathrm{dist}}

\newcommand{\rand}{\sim}

\newcommand{\eps}{\varepsilon}
\renewcommand{\epsilon}{\varepsilon}
\renewcommand{\subset}{\subseteq}

\newcommand{\FF}{\mathbb{F}}

\newcommand{\EE}{\mathbb{E}}

\newcommand{\mcB}{\mathcal{B}}
\newcommand{\mcC}{\mathcal{C}}
\newcommand{\mcD}{\mathcal{D}}
\newcommand{\mcE}{\mathcal{E}}

\newcommand{\mcH}{\mathcal{H}}

\newcommand{\mcM}{\mathcal{M}}

\newcommand{\mcV}{\mathcal{V}}

\newcommand{\BC}{\mathsf{BC}}
\newcommand{\HD}{\mathsf{HD}}
\newcommand{\ind}{\mathbf{1}}

\newcommand{\kalpha}{k_{m}}
\newcommand{\kbeta}{k_{u}}
\newcommand{\kgamma}{k_c}

\renewcommand{\Pr}{\mathbb{P}}
\renewcommand{\P}{\mathbb{P}}

\begin{document}
	
\title{From Algorithms to Connectivity and Back: \\ Finding a Giant Component in Random $k$-SAT}
\author{Zongchen Chen, Nitya Mani, and Ankur Moitra \thanks{\texttt{\{zongchen,nmani,moitra\}@mit.edu}}}
\affil{Department of Mathematics, Massachusetts Institute of Technology, Cambridge, MA 02139, USA}

\date{\today}
\maketitle

\begin{abstract}
We take an algorithmic approach to studying the solution space geometry of relatively sparse random and bounded degree $k$-CNFs for large $k$. In the course of doing so, we establish that with high probability, a random $k$-CNF $\Phi$ with $n$ variables and clause density $\alpha = m/n \lesssim 2^{k/6}$ has a giant component of solutions that are connected in a graph where solutions are adjacent if they have Hamming distance $O_k(\log n)$ and that a similar result holds for bounded degree $k$-CNFs at similar densities. We are also able to deduce looseness results for random and bounded degree $k$-CNFs in a similar regime.

Although our main motivation was understanding the geometry of the solution space, our methods have algorithmic implications. Towards that end, we construct an idealized block dynamics that samples solutions from a random $k$-CNF $\Phi$ with density $\alpha = m/n \lesssim 2^{k/52}$. We show this Markov chain can  with high probability be implemented in polynomial time and by leveraging spectral independence, we also observe that it mixes relatively fast, giving a polynomial time algorithm to with high probability sample a uniformly random solution to a random $k$-CNF. Our work suggests that the natural route to pinning down when a giant component exists is to develop sharper algorithms for sampling solutions to random $k$-CNFs.
\end{abstract}

\section{Introduction}

\subsection{Motivation and background}
We let $\Phi \sim \Phi_k(n, m)$ be a $k$-CNF formula on $n$ boolean variables with $m$ clauses, each drawn uniformly at random from the set of clauses of size $k \ge 3$. The solution space geometry of random $k$-CNFs exhibits a variety of threshold behaviors as the density $\alpha := m/n$ of the formula varies, which have been extensively studied by physicists, mathematicians, and computer scientists~\cite{BMW00,ZDE08,MPZ02,KMR07,ACH08,CP12,DSS22}.

\begin{figure}[ht!]
    \centering
    \includegraphics[width=0.9\linewidth]{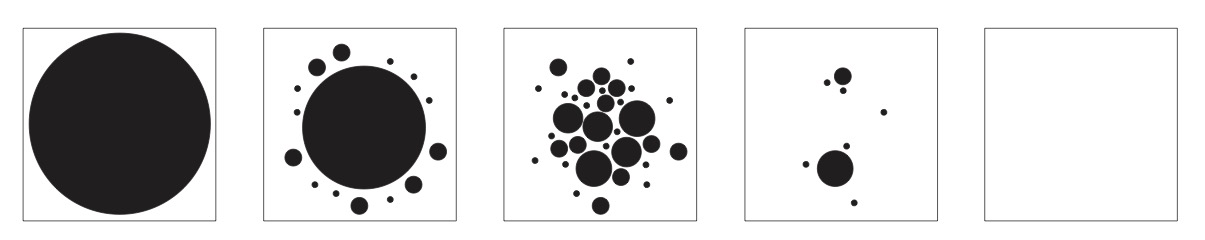}
    \caption{Heuristic phase diagrams such as above~\cite{KMR07} depict the predicted evolution of the structure of the solution space of a random $k$-CNF as the density $\alpha$ of the formula increases from left to right. We primarily study the leftmost regime.}
    \label{fig:solspace}
\end{figure}

Many involved heuristics in statistical physics make predictions about the geometry of the solution space of a random $k$-CNF instance, often depicted in diagrams like~\cref{fig:solspace}. Many phases and transitions in this diagram are precisely understood. 
For example, the \textit{satisfiability threshold} (pictured in the transition to the rightmost image in~\cref{fig:solspace}) was determined by~\cite{DSS22}; the satisfiability threshold characterizes the density at which a random $k$-CNF transitions from being satisfiable with high probability to being unsatisfiable with high probability. Another transition of interest is the \textit{clustering threshold,} above which the solution space of a random $k$-CNF shatters into exponentially many linearly separated connected components, each of which contains an exponentially small fraction of the satisfying assignments of the formula, as rigorously understood in~\cite{CP12,ACR11,MMZ05,MMZ205}.

In the lower-density regime, the solution space geometry of random $k$-CNFs appears poorly understood. It is widely believed that beneath a critical clause density, the solution space of a random $k$-CNF is ``connected.'' 
However, from the literature, it is not even clear what ``connected'' means. Connectivity is sometimes used in the statistical physics literature as a characterization of the entropy or energy profile of the solution space of a random $k$-CNF formula as in~\cite{ZDE08}. In such settings, connectivity is often characterized by an absence of clustering behavior, leaving somewhat of a mystery as to the graphical properties of the solution space of a low density random $k$-CNF.

Conjectures about connectivity take different forms, and different notions of what connectivity might mean are articulated in~\cite{ZDE08,KMR07,CP12}. The most common precise notion of connectivity is with respect to Hamming distance, i.e. understanding connectivity properties of the graph of solutions to a random $k$-CNF, where solutions are $f(n)$-connected if their Hamming distance is at most $f(n).$
At lower densities, random $k$-CNFs still can have isolated solutions far in Hamming distance from other satisfying assignments. However, the prevailing belief is that below some threshold, the overwhelming majority of solutions to a random $k$-CNF lie in a \textit{giant component} that is $o(n)$-connected.

Much more is known about related notions and local versions of connectivity, like \textit{looseness}, which characterizes how rigid a particular satisfying assignment is. Roughly speaking, a satisfying assignment to a formula is $f(n)$-loose if any variable can be flipped to yield a new satisfying assignment by changing at most $f(n)$ additional variable assignments.~\cite{ACH08}, the authors showed $o(n)$-looseness holds in the connectivity regime for related, simpler random models, random $q$-coloring, and hypergraph $2$-coloring,  conjecturing that $o(n)$-looseness holds for random $k$-CNF instances below the clustering threshold. This conjecture was partially resolved in~\cite{CP12}, where in an analysis of the \textit{decimation process} for random $k$-SAT, the authors observed that with high probability over formulae and satisfying assignments at least 99\% of the variables were $O(\log n)$-loose.

Looseness, however, is a local notion, not a global one. The set of elements in $\{0, 1\}^n$ that have Hamming weight at least $2n/3$ or at most $n/3$ is $1$-loose, but $\Omega(n)$-connected. Further, all previous approaches to establish looseness (see~\cref{s:prevwork} for details) proceed by taking steps in a small neighborhood of a satisfying assignment; thus, they cannot hope to connect arbitrary solutions potentially far away in the solution space. Consequently, a global, algorithmic perspective is needed to answer questions like the following:

\begin{qn}
In the connectivity regime, are solutions to a random $k$-CNF that lie in the giant component $O(\log n)$-connected? Or, is it necessary sometimes to change $\Omega(n^{1-\eps})$ variable assignments at a time to navigate between pairs of solutions on a path comprising only satisfying assignments?
\end{qn}

A primary contribution of our work is to take a global approach to understanding the solution space geometry of random $k$-CNFs by designing algorithms that sample from the solution space and take relatively short steps.

We take inspiration from the recent literature on sampling approximately uniformly random solutions to bounded degree CNFs in the Lov\'asz local lemma regime, as introduced in the deterministic counting approach of~\cite{MOI19} and further extended and strengthened in a variety of ways, such as in~\cite{FEN21,GAL19,JPV21}.

\subsection{Main results}
Throughout this work we study both bounded degree and random $k$-CNFs on variable set $V$ of size $n$ and with $m$ clauses, letting $\alpha = m/n$. We view $k$ as fixed but sufficiently large and let $n \rightarrow \infty$.

\begin{notn}
We employ the following further notation conventions.
\begin{itemize}
    \item We say $n$-variable $k$-CNF $\Phi$ is a \textit{$(k, d)$-formula} if $\Phi$ if  each variable appears in at most $d$ clauses of $\Phi$. 
    \item We say $n$-variable $k$-CNF $\Phi \sim \Phi_k(m, n)$ is a \textit{random $k$-CNF formula} if $\Phi$ consists of $m$ clauses, each of which is uniformly randomly sampled from the space of clauses of size $k$ on $n$ variables.
    \item Let $\Omega = \Omega_{\Phi}$ be the space of satisfying assignments of $k$-CNF $\Phi$.
    \item Given an assignment $\sigma$ to $\Phi$, we view $\sigma \in \{0, 1\}^n$ and let $\|\sigma\|_1$ be the number of variables $\sigma$ assigns to be $1$ or ``True''. Throughout, we implicitly consider variable assignments in $\FF_2^r$ so that $\|\cdot\|_1$ encodes Hamming weight.
\end{itemize}
\end{notn}

As discussed above, a primary goal of our work will be to develop an understanding of the connections between the solution space geometry of $k$-CNF $\Phi$ and algorithms for efficiently sampling from the solutions of $\Phi$. We will concern ourselves with the following precise notion of connectivity.

\begin{defn}[$D$-Connectivity]
We say a sequence of satisfying assignments $\zeta_0 \leftrightarrow \zeta_1 \leftrightarrow \cdots \leftrightarrow \zeta_{\ell}$ to some $k$-CNF $\Phi$ is a \textit{$D$-path} if $\|\zeta_{i} - \zeta_{i-1}\|_1 \le D$ for each $i \in [t]$. 
We say two satisfying assignments $\sigma, \sigma' \in \Omega_{\Phi}$ are \textit{$D$-connected} if there exists a $D$-path connecting $\sigma$ and $\sigma'$ (that is, $\zeta_0 = \sigma$ and $\zeta_\ell = \sigma'$).
\end{defn}

A unifying theme of previous approaches to counting and sampling CSP solutions is a tool called \textit{marking}, first introduced in~\cite{MOI19}, which uses projection from the original state space of satisfying assignments to a more well connected space of partial assignments that can be algorithmically lifted up to an appropriate satisfying assignment.

Marking-based deterministic and MCMC algorithms are mysterious at first glance, as they enable counting and sampling of $k$-CNF solutions even in regimes where the solution space is disconnected (i.e. not $1$-connected). This lack of connectivity stymies several natural, local algorithmic approaches, such as the classical Glauber dynamics. Some algorithms~\cite{FEN21,GAL19,JPV21} use rather complicated workarounds to tackle specific sampling tasks; however the ability of such methods to generalize is unclear (see~\cref{s:prevwork} for a longer discussion of previous work). The results in this work suggest that there is an underlying property that makes approximate counting and sampling possible in solution spaces not connected by Hamming distance $1$ steps. Our connectivity results hint that $O(\log n)$-connectivity of the solution space might be closer to the true algorithmic threshold; we use this idea as motivation for an algorithmic approach to connectivity that is ``natively $O(\log n)$-local.''

In this work, we leverage the idea of marking in a novel way to construct paths that certify global connectivity properties of the solution space of $k$-CNFs at densities close to where sampling algorithms are known. 
More precisely, we obtain the following results for bounded degree and random $k$-CNFs.

\begin{thm}[Connectivity: bounded degree]
\label{t:lllconnected} 
There exist constants $\gamma \ge 0.1742$ and $c > 0$ such that the following holds.
For every $(k, d)$-formula $\Phi$ with $d \le \frac{c}{k^3}2^{\gamma k}$, two satisfying assignments $\sigma$ and $\sigma'$ chosen uniformly at random are $O(dk^2 \log(n))$-connected with probability $1-o(1)$. 
In other words, a $1 - o(1)$ fraction of solutions in $\Omega$ are pairwise $O(dk^2 \log(n))$-connected.
\end{thm}

At a high level, above we use a \textit{marking} (a suitable choice of a subset of relatively low degree variables) to guide a process that enables us to traverse between two arbitrary satisfying assignments by making local, greedy updates to a single marked variable at a time and extending these updates to new satisfying assignment by changing $O_k(\log n)$ variables with high probability. Our process also enables us to deduce connectivity for random $k$-CNFs from connectivity results for bounded degree CNFs.

\begin{thm}[Connectivity: random]
\label{t:rand-connected}
For $\gamma > 0$ such that~\cref{t:lllconnected} holds for all $(k, d)$-formulae with $d \le \frac{c'}{k^3} 2^{\gamma k}$ for some $c' > 0$ and any $\zeta \in (0, 1)$, there exists constant $c = c(\gamma, \zeta) > 0$ such that the following holds.
With high probability over random $k$-CNF formula $\Phi \sim \Phi_k(m, n)$ such that $\alpha = m/n \le \frac{c}{k^3}2^{(1-\zeta)\gamma k}$, two satisfying assignments $\sigma$ and $\sigma'$ chosen uniformly at random are $O(\alpha k^6 \log(n))$-connected with probability $1-o(1)$. 
In other words, with high probability over $\Phi$, a $1 - o(1)$ fraction of solutions in $\Omega$ are pairwise $O(\alpha k^6 \log(n))$-connected.
\end{thm}

As noted, the above result gives a black-box reduction from a robust version of $O(dk^2 \log n)$ connectivity of a $1 - o(1)$-fraction of solutions in bounded degree $k$-CNFs to $O(\alpha k^6 \log n)$-connectivity of a $1 - o(1)$-fraction of solutions of a random $k$-CNF  of commensurate density (with high probability). The structural similarities we observe further suggest that improved bounds for sampling from bounded degree CNFs might yield corresponding improvements in connectivity for random $k$-CNFs. Further, our methods yield corresponding reductions from bounded degree CSPs to random CSPs in cases where random CSPs have few ``bad'' variables.

Our new applications of marking also have implications for other, more local, structural properties of the $k$-CNF solution space, like looseness.

\begin{defn}\label{d:loose}
Given $k$-CNF formula $\Phi$ and random solution $\sigma \rand \Omega$, variable $v \in V$ is $f(n)$-\textit{loose with respect to $\sigma$} if there exists $\tau \in \Omega$ with $\tau(v) \neq \sigma(v)$ and $\|\sigma - \tau\| \le f(n)$.

We say that $\Phi$ is \textit{$f(n)$-loose} if with high probability over $\sigma \in \Omega$, every variable is $f(n)$-loose with respect to $\sigma$.
\end{defn}

We observed earlier that looseness does not imply connectivity; in fact, the other direction of implication is also false as looseness is an incomparable goal to connectivity. Looseness requires that \textit{locally}, we are able to flip \textit{any} variable and get to a nearby solution rather than merely the existence of a path away from a solution. Nonetheless, we are able to deduce some nontrivial results about the looseness of the solution space of bounded degree and random $k$-CNFs. 

\begin{thm}[Looseness for $\Phi_k(m,n)$]\label{t:randloose}
Let $\zeta \in (0, \frac12)$ be an arbitrarily small constant.
Suppose that $\alpha = m/n \le 2^{(1 - \zeta) \gamma k}$ for constant $\gamma \ge 0.1742$ as in~\cref{t:lllconnected}. Then, with high probability $\Phi_k(m, n)$ is $O(\alpha k^6 \log n)$-loose. Thus, if $\alpha \le 2^{k/6}$, with high probability $\Phi_k(m, n)$ is $O(\alpha k^6 \log n)$-loose.
\end{thm}

We also observe a similar looseness result in a slightly denser regime for sufficiently sparse bounded degree $k$-CNFs in~\cref{t:lllloose}.

As the above conversation suggests, our results about connectivity and looseness are motivated by an algorithmic perspective. To that end, we give a polynomial time algorithm for approximately uniformly sampling a solution of a random $k$-CNF by showing that a natural block dynamics Markov chain on a projected state space mixes quickly. To exhibit fast mixing of this MCMC algorithm, we show that an associated projected distribution is $O(\log n)$-spectrally independent. This gives one of the first application of spectral independence to the task of sampling from random $k$-CNFs; spectral independence has been previously used to great effect as a tool in a variety of statistical physics sampling problems~\cite{AL20}, including sampling from random graphs of potentially unbounded degree in~\cite{BEZ21}. Using deterministic methods~\cite{GAL19} can an algorithm to approximately count solutions to a random $k$-CNF for $\alpha \lesssim 2^{k/300}.$

\begin{thm}\label{t:sampling}
Let $\Phi \sim \Phi_k(m, n)$ be a random $k$-CNF with $\alpha \le 2^{k/52}$. With high probability over $\Phi$, there exists a randomized polynomial time algorithm that with probability $1 - o(1)$ samples a uniformly random solution $\sigma \rand \Omega_{\Phi}.$ 
\end{thm}

Right before we posted our manuscript, we were informed that there are two new papers (independent, contemporaneous work) \cite{GGGH22+,HWY22+} which achieve nearly linear-time sampling algorithms for random $k$-CNF solutions with a better parameter regime.

\subsection{Main techniques}

Our connectivity results follow from designing and analyzing algorithms (\cref{alg:find-path,alg:rCNF-path}) that construct a path between two satisfying assignments to a bounded degree or random $k$-CNF. The path is fully contained in the solution space $\Omega_{\Phi}$ and moreover, adjacent pairs of solutions have Hamming distance $O_k(\log n)$ with high probability.

Both algorithms utilize a \textit{marking} $\mcM \subset V$ (see~\cref{s:marking}). A marking is a carefully chosen subset of variables that has a large intersection and non-intersection with every clause. One important property of our choice of marking is that even after conditioning on a large subset of marked variables, the remaining variables will satisfy a quantitative \textit{local uniformity} property (\cref{d:localuniform}).

Such a marking thus enables us to construct a two stage algorithm (\cref{alg:find-path}) to establish connectivity for bounded degree CNFs, that proceeds roughly as follows. Suppose that we are given a pair of solutions $(\sigma, \sigma')$. 
In the first stage, we sequentially update all marked variables of $\sigma$ so that they match the assignment of $\sigma'$. 
Now when we update one marked variable $v$ (e.g., flipping from True to False), we may need to flip some further subset of variables to ensure that we still have a satisfying assignment. We restrict ourselves to only flip unmarked variables (in addition to the chosen variable $v$) but \emph{not} other marked variables.
The fact that we can accomplish this without changing more than $O_k(\log n)$ variables also follows from having a good initial marking. 
After the first stage, we obtain a solution $\zeta$ satisfying $\zeta(\mcM) = \sigma'(\mcM) =: \tau'$.

In the second stage, we walk from $\zeta$ to $\sigma'$ by updating every connected component in the hypergraph associated to the formula we get when we simplify $\Phi$ using partial assignment $\tau'$ (we use local uniformity to show that these connected components have size $O_k(\log n)$ with high probability).
In this stage, we only update unmarked variables and every variable is updated at most once. By stitching these steps together, we obtain a path $\sigma \leftrightarrow \zeta \leftrightarrow \sigma'$.

As it turns out, by a more careful analysis of the above approach, one finds that $O_k(\log n)$-connectivity not only holds for pairs of random solutions from a bounded degree $k$-CNF with high probability but in a more general setting, where we draw solutions from a not-necessarily-uniform (but somewhat well-behaved) distribution, as precisely described in~\cref{t:robustconn}.
This observation is important in our extension to random $k$-CNFs. Unlike bounded degree CNFs, random CNFs can have variables of unbounded degree, complicating applications of the Lov\'asz local lemma and notions of local uniformity. However, at logarithmic scale, random CNFs behave much more like bounded degree CNFs. This can be made precise by defining the appropriate notion of a \textit{bad} variable $\mcV_{\bad}$ (see~\cref{d:goodbad}) and isolating these. 

To show connectivity in the random case, we construct in~\cref{alg:rCNF-path} a path of solutions between $\sigma, \sigma'$ that proceeds in two major steps. We first obtain a bounded degree $k$-CNF from $\Phi$ by conditioning on the partial assignment of $\sigma(\mcV_{\bad})$. We then choose $\psi,$ a uniformly random solution of a modification of $\Phi$ where we delete all bad variables and associated \textit{bad clauses} (which satisfies the bounded degree $k$-CNF we constructed from $\Phi$ via the partial assignment $\sigma(\mcV_{\bad})$). The robust notion of bounded degree connectivity of~\cref{t:robustconn} allows us to find a path of satisfying assignments between $\sigma(V \backslash \mcV_{\bad})$ and $\psi$ that we can lift to a path of satisfying assignments from $\sigma$ to $\psi_{\sigma} := \psi \cup \sigma(\mcV_{\bad})$ (here, robustness is important, since neither $\psi$ nor $\sigma(V \backslash \mcV_{\bad})$ are uniform samples from the set of satisfying assignments of the bounded degree CNF obtained by simplifying $\Phi$ under partial assignment $\sigma(\mcV_{\bad})$). We similarly construct a path of satisfying assignments between $\sigma'$ and some $\psi_{\sigma'}$. Using the fact that $\psi_{\sigma}, \psi_{\sigma'}$ agree except on bad variables and that connected components of bad variables are small (\cref{l:randbadcomp}), we are able to greedily construct a path between $\psi_{\sigma}, \psi_{\sigma'}$ and join these three paths together to establish connectivity

Looseness follows via similarly motivated arguments in both the random and bounded degree settings. In the case of random $k$-CNFs, we additionally derive a Markov chain-based polynomial time sampling algorithm by analyzing a block dynamics (\cref{a:randblockdynamics}) on the projected solution space of a random $k$-CNF onto a set of marked variables. By analyzing a delicate coupling (\cref{a:coupling}), we show that the marginal distribution of a uniformly random $k$-CNF solution on marked variables is $O_k(\log n)$ spectrally independent (see~\cref{l:si}), from which we can conclude fast mixing of the block dynamics. Using similar arguments to those used to establish that adjacent solutions in the connectivity paths constructed have distance $O_k(\log n),$ we are able to show that the steps of the block dynamics can be implemented efficiently with high probability (see~\cref{l:appxblock}).

Several of our techniques are motivated by related literature for sampling solutions of bounded degree $k$-CNFs and approximately counting solutions to random $k$-CNFs.

\subsection{Previous work}\label{s:prevwork}
\subsubsection{Connectivity}
There is considerable interest in understanding the phase diagram and transitions of the solution space of a random $k$-CNF. Physicists have developed sophisticated heuristics to explain the evolution of the solution space geometry of a random $k$-CNF and predict phase transitions in the associated graph~\cite{MPZ02,ZDE08,BMW00,KMR07}. We understand rigorously some aspects of this phase diagram very precisely, as in the landmark work of~\cite{DSS22},  characterizing the threshold for a $k$-CNF to be \textit{satisfiable} with high probability for large enough $k$. In low density regimes, there has been study of a \textit{clustering regime} where the solution space is (with high probability) comprised of exponentially many, linearly separated clusters, each of which contains an exponentially small fraction of the solutions to a random $k$-CNF. Several quantitative results describe the nature of such clusters, their Hamming distance separation, and associated \textit{frozen} variables in clusters, including~\cite{ACH08,CP12,ACR11,MMZ05,MMZ205}.

However, rigorous results about connectivity are missing; as noted earlier, even precise definitions of connectivity are often absent or inequivalent. Two common notions of connectivity are (a) a ``pure state'' entropic notion useful for numerical simulations and (b) connectivity of the solution space graphs where solutions are connected if their Hamming distance is $o(n)$ (see~\cite{CP12,ACH08,MPZ02}). These two notions are incomparable in general (as noted in~\cite{ZDE08}) and often even weaker notions of connectivity are applied, such as when connectivity is implicit defined as a failure of some specific clustering to appear (as in~\cite{KMR07}.

In this work, we pick a strong, robust notion of connectivity by showing that with high probability, the graph of the solution space of a sufficiently sparse random $k$-CNF, where solutions that have $O_k(\log n)$ Hamming distance are adjacent, has a giant connected component.

\subsubsection{Looseness}
Earlier work~\cite{ACH08} showed that random $k$-coloring and $k$-uniform hypergraph $2$-coloring instances are $o(n)$-loose up to the clustering threshold, and conjectured that the same held for random $k$-SAT instances. In~\cite{CP12}, the authors studied the decimation process and observed that up to the clustering threshold, a random $k$-SAT formula-assignment pair with high probability had the slightly different property: 99\% of variables are $O(\log n)$-loose.

Previous work showing the failure of $o(n)$-connectivity of the random $k$-SAT solution space above the clustering threshold shows that not only does connectivity fail, but so does looseness, with the solution space shattering into exponentially many, exponentially small clusters that are linearly separated with a linear fraction of \textit{frozen variables}~\cite{ACH08}. The converse result, that \textit{all} variables are $o(n)$-loose in a random $k$-SAT formula with high probability over formula and satisfying assignment is not known in as strong a sense.

\subsubsection{Sampling}
Approximately counting and sampling from a solution space of exponential size are fundamental problems in computing.
One of the most important solution spaces in computer science is the set of satisfying assignments to a given or random $k$-CNF formula. Since the solution space is not connected by Hamming distance $1$ moves~\cite{WIG19}, there is a barrier to the most naive Markov chain Monte Carlo approach. 

In~\cite{MOI19}, Moitra introduced \textit{marking} in the course of giving an algorithm to estimate the number of satisfying assignments of $k$-SAT for bounded degree $k$-CNFs, by leveraging the structure accorded to the variables by the Lov\'asz local lemma. This breakthrough gave a polynomial time algorithm, running in time $n^{O(d^2k^2)}$ for maximum degree $d$ $k$-CNFs provided $d \ll 2^{k/60}.$ . This work has since been improved to fixed-parameter tractable sampling and enumeration of $k$-CNF solutions in denser regimes.

In~\cite{FEN21}, the authors introduced an MCMC-based approach to sampling in the Lov\'asz local lemma regime, by applying a Markov chain to a connected projection of the original state space. These ideas have been extended to a more general \textit{state compression} method that applied to more general atomic CSPs in further work~\cite{JPV21}. Further, the original bound of~\cite{MOI19} on the maximum degree was improved to $d \lesssim 2^{k/20}$ by~\cite{FEN21} and to $d \lesssim 2^{k/5.741}$ by~\cite{JPV21}. In the setting of more general atomic CSPs in the LLL regime,~\cite{HE21} gave a perfect sample.

In more constrained settings, much further progress has been made. One such example are monotone $k$-CNF formulae, whose underlying solution space is connected. In~\cite{HSZ19}, the authors observed that efficient randomized algorithms exist for monotone $k$-CNF formulae of maximum degree $d \le c 2^{k/2}$ for some constant $c > 0$, which is sharp up to the choice of $c$ due to corresponding hardness results~\cite{BGGGS19}. They also observed that such algorithms also work in the setting of random regular monotone $k$-CNF formulae of degree $d \le c 2^k/k$.

Much less is known for more general random $k$-CNFs, which can have variables with unbounded degree. In~\cite{GAL19}, the authors adapted the marking approach of~\cite{MOI19} to give a polynomial time (for fixed average degree $d$ and $k$) algorithm for estimating the number of solutions to a random $k$-CNF, provided that $d \lesssim 2^{k/300}$. 
In a slightly different setting~\cite{BEZ21} used spectral independence to relax the bounded degree assumption to develop algorithms based on the Glauber dynamics for sampling $2$-spin systems of random graphs $G(n, d/n)$ in time $n^{1 + o(1)}$ for all $d$.

\subsection{Open questions}
Our approach suggests the heuristic that the connectivity threshold is closely tied to the efficacy of algorithms that traverse the solution space. This motivates using statistical physics heuristics to make precise conjectures about connectivity and associated conjectures about algorithmic tractability.

\begin{conj}
For $\alpha < (1 - o(1)) \frac{2^k}{k} \ln k,$ with high probability a random $k$-CNF $\Phi \sim \Phi_k(m, n)$ has a $1 - o(1)$ fraction of solutions in a $O_{k}(\log n)$-connected component. 
\end{conj}

\begin{conj}\label{c:loose}
For $\alpha < (1 - o(1)) \frac{2^k}{k} \ln k,$ with high probability a random $k$-CNF $\Phi \sim \Phi_k(m, n)$ has a $1 - o(1)$ fraction of solutions that are $O_k(\log n)$-loose. 
\end{conj}

\begin{conj}
For CSPs $\Psi$ with solution space is $O(\log n)$-connected under the Hamming distance, there exists a randomized polynomial-type algorithm to approximately sample a uniformly random solution to $\Psi.$
\end{conj}

\subsection{Organization}
In~\cref{sec:bounded-CNF}, we study bounded degree $k$-CNFs, beginning with some preliminaries before moving on to prove~\cref{t:lllconnected} and~\cref{t:lllloose}. In~\cref{sec:random-CNF}, we study connectivity, looseness, and sampling in random $k$-CNFs, beginning by proving~\cref{t:rand-connected,t:randloose}. We define a block dynamics Markov chain in~\cref{a:randblockdynamics}. We use this chain to prove~\cref{t:sampling}; we first give spectral independence bounds via a delicate coupling and then conclude fast mixing as a consequence. Due to the small size of the components we need to search over to compute the relevant marginal probabilities in the chain, the chain gives a polynomial time algorithm with high probability for approximate sampling. 
We also defer some proofs of technical details to~\cref{a:badcomps,a:missingsec3}.

\subsection*{Acknowledgements}
N.M. was supported by an Hertz Graduate Fellowship and the NSF GRFP. A.M. was supported by a Microsoft Trustworthy AI Grant, NSF Large CCF1565235, NSF CCF1918421 and a David and Lucile Packard Fellowship.

\section{Bounded degree CNFs}
\label{sec:bounded-CNF}

\subsection{Preliminaries}
Suppose that $\Phi$ is a fixed $(k, d)$-formula.
Let $\mu$ denote the uniform distribution on satisfying assignments to $\Phi$, i.e. the uniform distribution on $\Omega$. 

\begin{defn}
Given $(k, d)$-formula $\Phi$, let $H_{\Phi} = (V, \mcE)$ be the \textit{dependency (multi)hypergraph} where $V$ is the set of variables and $\mcE = \{ \text{var}(C) : C \in \Phi \}$ is the collection of clauses of $\Phi$ viewed as $k$-sets of variables.
\end{defn}

\begin{defn}
Let $\Phi$ be a $k$-CNF.
Given a partial assignment $X$ on $\Lambda \subset V$, let $\Phi^X$ be the result of simplifying $\Phi$ under $X$, so that $\Phi^X = (V^X, C^X)$ where $V^X = V \setminus \Lambda$ and $C^X$ is obtained from $C$ by removing all clauses that have been satisfied under $X$ and removing any appearance of variables in $\Lambda$. We let $H_{\Phi}^X$ be the associated (not necessarily $k$-uniform) hypergraph to $\Phi^X$ and for variable $v \in V \setminus X$, let $\mcE_v^X$ denote the connected component of $H_{\Phi}^X$ to which $v$ belongs.
\end{defn}

\begin{defn}
For an arbitrary set of variables $S \subset V$, let $\mu_S$ be the marginal distribution on $S$ induced by $\mu$, so formally
$$\forall \sigma \in \{0, 1\}^S, \quad \mu_S(\sigma) = \sum_{\tau \in \{0, 1\}^n : \tau|_S = \sigma} \mu(\tau)$$
Further, given some partial assignment $X \in \{0, 1\}^{\Lambda}$ for $\Lambda \subset V$ if $S \cap \Lambda = \emptyset$, we let $\mu_{S}^X(\cdot) := \mu_S(\cdot \mid X)$ be the marginal distribution on $S$ conditioned on the partial assignment on $\Lambda$ in $X$.
\end{defn}

\subsection{\texorpdfstring{$O(\log n)$}{O(log n)}-connectivity of \texorpdfstring{$k$}{k}-CNF in the local lemma regime}
\label{s:boundcon}

In this section, we show that the solution space of a bounded-degree $k$-CNF has a $O(dk^2 \log n)$-connected giant component in the local lemma regime, proving~\cref{t:lllconnected}. 

Our main result shows that two uniformly random solutions of a bounded-degree $k$-CNF in the local lemma regime are $D$-connected for $D = O(dk^2 \log n)$ with high probability, which implies the existence of a giant $O(dk^2\log n)$-connected component consisting of almost all solutions. 
In fact, we will show a more robust result (see~\cref{t:robustconn}) that will enable us to deduce connectivity for random $k$-CNFs in a black box manner for a similar density regime.

Our proofs of~\cref{t:lllconnected,t:robustconn} utilize a \textit{marking} of variables. Previously, the marking technique has been successfully applied to give fast samplers for the uniform distribution on all solutions and efficient counting algorithms for the number of solutions. 
We give new applications of marking and establish properties of solution space geometry which were not known before, including connectivity and looseness.

\subsubsection{Markings}\label{s:marking}

We require the following version of a marking of variables.
\begin{lemma}[{\cite[Proposition 3.3 (2)]{JPV21}}]\label{l:alglll}
There exist constants $\gamma \ge 0.1742$ and $c > 0$ such that the following holds. 
For every $(k, d)$-formula $\Phi$ with $d \le  \frac{c}{k^3} 2^{\gamma k}$, 
there exists a set of marked variables $\mcM \subseteq V$ such that every clause has at least $\kalpha := 2\gamma k$ marked and $\kbeta := \gamma k$ unmarked variables.
Furthermore, there is a randomized algorithm that finds such a set of marked variables in polynomial time with probability $3/4$.
\end{lemma}

We will use a slightly more robust marking result that applies to CNFs that can have different numbers of variables in each clause.
\begin{defn}
$\Phi$ is a \textit{$(k, \zeta, d)$-CNF} if $\Phi$ is a CNF on $n$ variables such that each variable appears in at most $d$ clauses of $\Phi$ and each clause has between $(1-\zeta)k$ and $k$ literals.
\end{defn}

\begin{lem}[Marking]
\label{lem:marking-robust}
There exist constants $\gamma \ge 0.1742$, $c > 0$, and $k_0>0$ such that the following holds for any integer $k \ge k_0$ and any $\zeta \in [0,1/2)$. 
Let $\Phi$ be a $(k,\zeta,d)$-CNF with $d \le \frac{c}{k^3} 2^{\gamma(1-\zeta) k}$. 
Then there exists a set $\mcM \subseteq V$ of marked variables, such that each clause has at least $k_m = 2\gamma(1-\zeta)k$ marked variables and at least $k_u = \gamma(1-\zeta)k$ unmarked variables. 
\end{lem}

Note that for the applications in this section we only need the existence of such a marking but we do not actually need a polynomial time algorithm for finding it.

\begin{proof}
For $\gamma = 0.1742$, we have that $d \lesssim \frac{2^{\gamma(1 - \zeta)k}}{k^3}$ and thus, for a $(k, \zeta, d)$-CNF $\Phi$, we can fix a marking $\mcM$ per~\cref{l:alglll} where every clause has at least $\kalpha = 2 \gamma(1-\zeta) k$ marked and $\kbeta = \gamma(1-\zeta) k$ unmarked variables.
\end{proof}

\begin{rem}
Note that in the above parameter regime, $2^{\kbeta} \ge 2 e d k^2, 2^{\kalpha} \ge 4 e^2 d^2 k^2$.
\end{rem}

Throughout the remainder of this section, we fix some marking $\mcM \subseteq V$ that satisfies the conditions of \cref{lem:marking-robust}.
Such a marking allows us to get an approximate ``uniformity'' of satisfying assignment restrictions to subsets of marked variables, which is a consequence of the following corollary of the Lov\'asz Local Lemma (c.f Corollary 2.2 in~\cite{FEN21}).

\begin{prop}\label{p:lll}
Given a CNF $\Phi = (V, C)$ such that each clause contains at least $k_1$ variables and at most $k_2$ variables and each variable belongs to at most $d$ clauses. For any $s \ge k_2$, if $2^{k_1} \ge 2e d s$, then there exists a satisfying assignment for $\Phi$, and for any $v \in V$
$$\max_{c \in \{0, 1\}} \mu[X(v) = c] \le \frac12 \exp\left( \frac{1}{s}\right),$$
where $X$ is a uniformly random satisfying assignment.
\end{prop}

The above result implies that while $\mu$ may not induce the uniform distribution on each variable's assignment, it assigns probabilities far from $0$ to each possible True/False assignment, a phenomenon we capture in the following notion of being \textit{locally uniform.}

\begin{defn}[Local uniformity]\label{d:localuniform}
We say that a distribution $\pi$ on $\{0, 1\}^{\mcM}$ is $s$-\textit{locally uniform} if for all $U \subseteq \mcM$ and all assignments $\tau \in \{0,1\}^U$ on $U$, one has
\[
\pi(X(U) = \tau) \le \frac{1}{2^{|U|}} \exp\left( \frac{|U|}{s} \right).
\]
\end{defn}

We can show the following local uniformity result. 

\begin{lemma}\label{l:vtxapproxuniform}
Let $\Phi$ be a $(k, \zeta, d)$ formula with marking per~\cref{lem:marking-robust} such that $2^{\kbeta} \ge 2 e d s$ for some $s \ge k$.
Fix any partial assignment $X$ on $S \subset \mcM$. Then, for any $v \in V \setminus U$ and $a \in \{0, 1\}$,
$$0<\mu_v(a \mid X) \le \frac12 \exp \left( \frac{1}{s} \right).$$
In particular, the marginal distribution $\mu_\mcM$ on marked variables $\mcM$ is $k$-locally uniform.
\end{lemma}
\begin{proof}
Let $\Phi^X$ be the CNF formula obtained by deleting all the clauses satisfied by $X$ and the variables in $S$. Let $\mu^X$ be the uniform distribution on satisfying assignments of $\Phi^X$, noting that $\mu^X_v(\cdot) = \mu_v(\cdot \mid X)$. Since $S \subset \mcM$, every clause left in $\Phi^X$ contains between $\kbeta$ and $k$ variables, each of which belongs to at most $d$ clauses. Since $2^{\kbeta} \ge 2eds$, by \cref{p:lll}, we have that for $c \in \{0, 1\}$,
$$\mu_v^X(c) = \mu_v(c \mid X) \le \frac12 \exp \left( \frac{1}{s} \right).$$
Thus, $\mu_\mcM$ is $k$-locally uniform.
\end{proof}

With this language, we can now state a more robust version of~\cref{t:lllconnected} that we prove.

\begin{thm}[Connectivity]
\label{t:robustconn}
There exist constants $\gamma \ge 0.1742$, $c > 0$, and $k_0>0$ such that the following holds for any integer $k \ge k_0$ and any $\zeta \in [0,1/2)$. 
Let $\Phi$ be a $(k, \zeta, d)$-CNF with $d \le \frac{c}{k^3} 2^{\gamma(1-\zeta) k}$. 
Suppose that $\pi$ and $\pi'$ are distributions over solutions of $\Phi$ whose marginal distributions on marked variables $\mcM$ are both $k$-locally uniform. 
If $\sigma,\sigma'$ are two independent random assignments chosen from $\pi,\pi'$ respectively, then whp $\sigma$ and $\sigma'$ are $O(dk^2\log n)$-connected.
\end{thm}

Assuming the above,~\cref{t:lllconnected} immediately follows.

\begin{proof}[Proof of~\cref{t:lllconnected}]
The theorem follows from \cref{t:robustconn} once we take $\zeta = 0$ in \cref{t:robustconn}. 
Note that the existence of such a marking $\mcM$ is guaranteed by \cref{l:alglll}, 
and the $k$-local uniformity of $\mu_{\mcM}$ is shown by \cref{l:vtxapproxuniform}.
\end{proof}

Thus, it remains to show~\cref{t:robustconn}.

\subsubsection{Algorithm}

We consider~\cref{alg:find-path} that receives two satisfying assignments of $(k, d)$-CNF $\Phi$ as the input and constructs a path between them. 

\begin{algorithm}[t]
\caption{Finding an $O(kd \log n)$-path between two solutions}\label{alg:find-path}
\KwIn{a CNF $\Phi$, a set $\mcM = \{v_1, \ldots, v_{\ell}\}$ of marked variables, two solutions $\sigma,\sigma'$} 

\medskip
$\zeta_0 = \sigma$\;
\medskip

\Comment*[h]{Stage 1: Update marked variables}\;
\For{$i \in [\ell]$}{
  Find $\zeta_i \in \Omega$ with marked variables specified by
  $
  \zeta_i (v_j) = 
  \begin{cases}
  \sigma'(v_j), & j \le i\\
  \sigma(v_j), & j > i
  \end{cases}
  $
  such that $\|\zeta_i - \zeta_{i-1}\|_{1}$ is minimized\; \label{alg:step2b}
}

\medskip
$\xi_0 = \zeta_\ell$\;
\medskip

\Comment*[h]{Stage 2: Update unmarked variables}\;
Let $\tau' = \sigma'(\mcM)$
and suppose that $H_{\Phi}^{\tau'}$ has connected components $\mcE_1, \ldots, \mcE_r$\; 
\For{$i \in [r]$}{ 
	Let $\xi_i \in \Omega$ be defined as
	$
	\xi_i(v) = 
	\begin{cases}
	\sigma'(v), & v \in \left(V \setminus \bigcup_{j = 1}^r \mcE_j\right) \cup \left(\bigcup_{j = 1}^{i} \mcE_j\right) \\
	\sigma(v), & v \in \left(\bigcup_{j = i+1}^{r} \mcE_j\right)
	\end{cases}
	$\;
	\label{alg:step3}
}

\medskip
\KwOut{The path $\sigma = \zeta_0 \leftrightarrow \cdots \leftrightarrow \zeta_\ell = \xi_0 \leftrightarrow \cdots \leftrightarrow \xi_r = \sigma'$}
\end{algorithm}
To prove \cref{t:robustconn}, it suffices to show that the output of \cref{alg:find-path} is with high probability a $D$-path in the solution space for $D = O(dk^2 \log n)$ when the inputs $\sigma \sim \pi$ and $\sigma' \sim \pi'$ arise from distributions with $k$-locally uniform marginals on $\mcM$.

We need the following two lemmas to establish this fact.
The first lemma shows that all the truth assignments $\zeta_i$, $\xi_i$ in the algorithm exist and satisfy the formula  (i.e. the algorithm is well-defined), implying our constructed path is indeed a valid path comprising only satisfying assignments..

\begin{lemma}\label{lem:I}
\cref{alg:find-path} is well-defined in the following sense:
\begin{enumerate}
\item \label{item:I-1} It is always possible to implement Line~\ref{alg:step2b} such that $\zeta_i \in \Omega$.
\item \label{item:I-2} We have $\xi_i \in \Omega$ for each $i \in [r]$.
\end{enumerate}
\end{lemma}

The second lemma shows that whp, two adjacent assignments differ by at most $O(dk^2\log n)$ variables, which builds upon the properties of our marking.

\begin{lemma}\label{lem:IV}
With high probability over the choices of $\sigma \sim \pi, \sigma' \sim \pi'$ such that $\pi, \pi'$ induce $k$-locally uniform marginal distributions on $\mcM$, the following pair of upper bounds holds:
\begin{enumerate}
\item $\|\zeta_i - \zeta_{i-1}\|_1 = O(dk^2 \log n)$ for all $i \in [\ell]$.
\item $\|\xi_i - \xi_{i-1}\|_1 = O(dk^2 \log n)$ for all $i \in [r]$.
\end{enumerate}
\end{lemma}

We first present the proof of \cref{t:robustconn}, and then in \cref{subsec:pf-I-IV} we prove \cref{lem:I,lem:IV}.

\begin{proof}[Proof of \cref{t:robustconn}]
Take input bounded degree CNF $\Phi$ and associated marking $\mcM \subseteq V$ as in \cref{lem:marking-robust}. 
With high probability over the choice of two random solutions $\sigma \sim \pi$ and $\sigma' \sim \pi'$ such that the marginals of $\pi, \pi'$ on $\mcM$ are $k$-locally uniform, the output path of \cref{alg:find-path} is well-defined by \cref{lem:I} and satisfies that $\|\zeta_i - \zeta_{i-1}\|_1 = O(dk^2 \log n)$ for all $i \in [\ell]$ and $\|\xi_i - \xi_{i-1}\|_1 = O(dk \log n)$ for all $i \in [r]$ by \cref{lem:IV}.
Hence, it is a $D$-path in the solution space $\Omega$ for $D = O(dk^2 \log n)$ as we wanted.
\end{proof}

\subsubsection{Proof of~\texorpdfstring{\cref{lem:I,lem:IV}}{Lemmas 2.12 and 2.13}}
\label{subsec:pf-I-IV}

We first prove \cref{lem:I} which is relatively simple. 
\begin{proof}[Proof of~\cref{lem:I}]
For \cref{item:I-1}, we deduce from \cref{p:lll} that for any $v \in V$ and partial assignment $X \in \{0, 1\}^{\mcM \setminus \{v\}}$ or $X \in \{0, 1\}^{\mcM}$, we can extend $X$ to some satisfying assignment $\sigma \in \Omega$.

Next consider \cref{item:I-2}. Fix a formula $\Phi$. All clauses that do not appear in $H_{\Phi}^{\tau'}$ were satisfied by partial assignment $\tau'$. Now consider two satisfying assignments $\rho, \rho'$ such that $\rho(\mcM) = \rho'(\mcM) = \tau'$. Let $H_{\Phi}^{\tau'}$ have connected components $\mcE_1, \ldots, \mcE_r$. In particular, $\rho|_{\mcE_i}$ and $\rho'|_{\mcE_i}$ each satisfy all clauses incident to any variable in $\mcE_i$. Each clause remaining in $\Phi^{\tau'}$ is associated to exactly one $\mcE_i$. Consequently, any mixture assignment $X$ such that $X|_{\mcM} = \tau'$, and $X|_{\mcE_i} \in \{\rho_{\mcE_i}, \rho'_{\mcE_i}\}$ chosen arbitrarily for each $i \in [r]$ yields a satisfying assignment (any variables that do not appear in $\mcM \cup \bigcup_{i = 1}^r \mcE_i$ can be chosen arbitrarily). 
This shows \cref{item:I-2}. 
\end{proof}

The rest of this subsection is devoted to the proof of \cref{lem:IV}.
We will deduce \cref{lem:IV} from the observation that (with high probability over our initial choice of $\sigma, \sigma'$), given any partial assignment $X \in \{0, 1\}^{\mcM}$ that arises in the algorithm, at each intermediate step of the algorithm, the connected components of $H_{\Phi}^X$ have size $O(dk \log n).$
To prove this, we need the following local uniformity lemma, which crucially holds for any assignment in the path constructed by \cref{alg:find-path}.

\begin{lemma}
\label{l:alguniform}
Suppose $2^{\kbeta} \ge 2 e d k$.
Consider $\zeta_i \in \Omega$ and restriction to $\mcM$, $\tau^{(i)}$ for $i \in [\ell]$ that arises in Line~\ref{alg:step2b} of \cref{alg:find-path} when the algorithm is initialized with two random satisfying assignments $\sigma, \sigma' \in \Omega$ from $\pi,\pi'$ respectively.
Then the distribution of $\tau^{(i)}$ is $k$-locally uniform. 
\end{lemma}

Consider the execution of Line~\ref{alg:step2b} of \cref{alg:find-path}. Recall that $\mcM = \{v_1, \ldots, v_{\ell}\}$ and fix $i \in [\ell+1]$. Let $X = \tau^{(i-1)} \cap \tau^{(i)}$ be the assignment on $\mcM \setminus \{v_i\}$ common to these two partial assignments for $i \le \ell$ and let $X = \sigma'(\mcM)$ if $i = \ell + 1$. 
Let $\mcB_i$ denote the bad event that there is some connected component $\mcE_j^X$ of $H_{\Phi^X}$ such that $|\mcE_j^X| > 4dk \log n$. 
Then we can show from the local uniformity \cref{l:alguniform} that with high probability none of the bad events $\mcB_i$ will happen.

\begin{lemma}\label{l:smallcomp}
Suppose $2^{\kalpha} \ge 4 e^2 d^2 k^2$ and $2^{\kbeta} \ge 2 e d k$. 
Let $\mcM' \subseteq \mcM$ such that either $\mcM' = \mcM \setminus \{v\}$ for some $v \in \mcM$ or $\mcM' = \mcM$. 
For an assignment $X$ on $\mcM'$, let $\mcB$ denote the bad event that there is some connected component $\mcE_j^X$ of $H_{\Phi^X}$ such that $|\mcE_j^X| > 4dk \log n$. 
If $\pi$ is a distribution on satisfying assignments whose marginal on $\mcM$ is $k$-locally uniform, then $\P(\mcB) \le 1/n^2$ where $X$ is chosen from $\pi_{\mcM'}$. 
In particular, for each $i \in [\ell+1],$ in \cref{alg:find-path}, $\P[\mcB_i] \le \frac{1}{n^2}$. 
\end{lemma}

We are now ready to prove \cref{lem:IV} assuming the previous pair of lemmas.

\begin{proof}[Proof of~\cref{lem:IV}]
(1) Let $v_i$ be the $i$th marked vertex encountered by \cref{alg:find-path}. Let $X = \tau^{(i-1)} \cap \tau^{(i)}$ denote the shared assignment on $\mcM \setminus \{v_i\}$ of $\zeta_i$ and $\zeta_{i-1}$, and observe that $\zeta_{i-1} \in \Omega$. Then, the claim follows from the observation that the connected component $\mcE_j^X \subset H_{\Phi^X}$ containing $v_i$ has size $O(dk \log n)$ with high probability (per \cref{l:smallcomp}), and thus we can update $\zeta_{i-1}$ to a satisfying assignment $\zeta_i$ that restricts to $\tau^{(i)}$ on $\mcM$, by only updating a subset of the $O(dk \log n)$ variables in $\mcE_j^X$ after updating $v_i$'s assignment to be $\sigma_2(v_i)$.

(2) Analogously, let $X = \tau' = \sigma'|_{\mcM}$ be the assignment on $\mcM$ shared by $\xi_0$ and $\xi_r = \sigma'$. By \cref{l:smallcomp} with probability $1 - O(1/n^2)$, all of the connected components of $H_{\Phi}^X$, $\mcE_1^X, \ldots, \mcE_r^X$ have size $O(dk \log n)$. Then by construction, Line~\ref{alg:step3} in \cref{alg:find-path} updates $\le |\mcE_j^X| = O(dk \log n)$ vertices between $\xi_{i-1}$ and $\xi_i$ for each $i \in [r]$. Since $\xi_0, \xi_r \in \Omega$, these intermediate assignments are all satisfying assignments since variables in each connected component can be independently assigned to satisfy a suite of disjoint clauses. This gives the desired result.
\end{proof}

\subsubsection{Proof of~\texorpdfstring{\cref{l:alguniform,l:smallcomp}}{Lemmas 2.14 and 2.15}} 
\label{subsec:uniformity}

We prove~\cref{l:alguniform}, which establishes some form of local uniformity for a random solution. 

\begin{proof}[Proof of \cref{l:alguniform}]
We need to show that for any partial assignment $\tau$ on $U \subseteq \mcM$ it holds that
$$\P[\zeta_i(U) = \tau] = \P[\tau^{(i)}(U) = \tau] \le \frac{1}{2^{|U|}} \exp\left(\frac{|U|}{k}\right).$$
Let $\mcM = \{v_1, \ldots, v_{\ell}\}$ be the ordering that the algorithm processes marked vertices and let $U = \{v_{i_1}, \ldots, v_{i_L}\}$ with $i_1 < i_2 < \cdots < i_L$ for some $L \le \ell$. Per \cref{l:vtxapproxuniform}, for any $j \in [L]$, $c \in \{0, 1\}$, and partial assignment $X' \in \{0, 1\}^{\mcM \setminus v_{i_j}}$ we have 
\begin{equation}\label{e:ineq}
\mu_{v_{i_j}}(c \mid X') \le \frac12 \exp\left( \frac{1}{k} \right).
\end{equation}
Let $t_{v_{i_j}} = i_j$ if $i_j \le i$ else let $t_{v_{i_j}} = 0$. 
By the chain rule,
\begin{align*}
\P[\tau^{(i)}(U) = \tau] &= \prod_{j = 1}^{L} \P[\tau^{(i)}(v_{i_j}) = \tau(v_{i_j}) \mid \tau^{(i)}(v_{i_p}) = \tau(v_{i_p}),\,1\le p < j] \\
&= \prod_{j = 1}^{L} \P[\tau^{(i)}(v_{i_j}) = \tau(v_{i_j}) \mid \tau^{(t_{v_{i_p}})}(v_{i_p}) = \tau(v_{i_p}),\,1\le p < j] \\
&\le \left(\frac12 \exp \left( \frac{1}{k} \right) \right)^{L},
\end{align*}
where the last inequality holds per Inequality~\eqref{e:ineq} and by the fact that $\sigma \sim \pi, \sigma' \sim \pi'$ and $\pi,\pi'$ have marginal distributions on $\mcM$ that are $k$-locally uniform (and thus we can apply \cref{l:vtxapproxuniform}). 
\end{proof}

The rest of this subsection is devoted to the proof of \cref{l:smallcomp}. 
We begin by recalling the useful notion of a $2$-tree and some relevant results about $2$-trees in $k$-uniform hypergraphs.

\begin{defn}[Line Graph]
Given a hypergraph $H = (V, E)$, the \textit{line graph} of $H$, denoted $\Lin(H)$ is the hypergraph with vertex set $V(\Lin(H)) = E(H)$ and where $(e_1, e_2) \in E(\Lin(H))$ if $e_1 \cap e_2 \neq \emptyset$, i.e. if edges $e_1, e_2$ share a vertex.
\end{defn}

\begin{defn}[$2$-Tree]
\label{def:2-tree}
Given a graph $G = (V, E)$, a subset of vertices $T \subseteq V$ is called a \textit{2-tree} if no pair $u, v \in T$ are adjacent (i.e. $\dist_G(u, v) \ge 2$ for all $u, v \in T$) and if adding an edge between every $u, v \in T$ that have $\dist_G(u, v) = 2$ makes $T$ connected.
\end{defn}

\begin{lem}[\cite{BCKL13}]\label{o:2tree}
Let $G = (V, E)$ be a graph with maximum degree $\Delta(G)$. Fix a vertex $v \in V$. The number of $2$-trees in $G$ of size $\ell$ containing $v$ is at most $\frac12 ((e \Delta(G))^2)^{\ell - 1}$.
\end{lem}

\begin{lemma}[{\cite[Lemma 5.8]{FEN21}}]\label{l:2treelin}
Let $H = (V, \mcE)$ be a $k$-uniform hypergraph such that each vertex belongs to at most $d$ hyperedges. Let $B \subseteq \mcE$ be a subsets of hyperedges that induces a connected subgraph in $\Lin(H)$ and let $e \in \mcB$ be an arbitrary hyperedge. Then, there must exist a $2$-tree $T \subset B$ in the graph $\Lin(H)$ such that $e \in T$ and $|T| = \lfloor \frac{|B|}{kd} \rfloor$.
\end{lemma}

Similarly to Lemma 5.3 of~\cite{FEN21}, we proceed by considering $H_{\Phi}^X$ for a partial assignment $X$ that yields a large bad component.We construct a $2$-tree in $H_{\Phi}^X$ that will allow us to show that when $\mcB_i$ occurs, there are many independent unlikely events, related to failures of local uniformity, that must \textit{all} occur. This will imply that $\mcB_i$ is very unlikely.

\begin{proof}[Proof of \cref{l:smallcomp}]
Fix $i \in [\ell+1]$ where $\mcM = \{v_1, \ldots, v_{\ell}\}$ and consider associated $X = \tau^{(i-1)} \cap \tau^{(i)}$, the assignment on $\Lambda = \mcM \setminus \{v_i\}$ common to these two partial assignments. Observe that bad event $\mcB_i$ occurs if some connected component of $H_{\Phi^X}$, $\mcE_j^X$ has $|\mcE_j^X| > 4dk \log n$.

Given some $e \in \mcE$, where $H_{\Phi} = (V, \mcE)$, let $\mcB_e$ denote the bad event that both $e \in \mcE^X$ (i.e. clause $e$ remains unsatisfied under $X$) and that $|\mcE_e| \ge 4 dk \log n$, where $\mcE_e$ is the connected component of $H_{\Phi^X}$ containing the vertices of $e$. By a union bound, observe that 
$$\P[\mcB_i] \le \P\left[\bigcup_{e \in \mcE} \mcB_e\right] \le \sum_{e \in \mcE} \P[\mcB_e].$$
Thus, our goal will be to give a good bound on $\P[\mcB_e]$. If $\mcB_e$ occurs, there is a subset $B$ such that $e \in B \subset \mcE$ with $|B| = \lceil 4dk \log n\rceil =: L$ such that $B$ is connected in $\Lin(H)$ and all hyperedges in $B$ are bad (unsatisfied by partial assignment $X$). Letting $r = \lfloor \frac{L}{dk} \rfloor$ by \cref{l:2treelin}, this implies we can find a $2$-tree $T \subset B$ with $e \in T$ of size $|T| = r$.

Observe that for every edge $e \in \mcE$, $|e \cap \Lambda| \ge \kalpha - 1$ and that the hyperedges in $T$ are disjoint. Since we assumed that $2^{\kbeta} \ge 2 e d k$, we apply \cref{l:alguniform} to observe that 
\begin{align*}
\P[\text{all hyperedges in }T\text{ bad}] &= \P[\text{all marked variables in }T\text{ unsatisfied}] \\
&\le \left(\frac12\right)^{(\kalpha-1)r} \exp\left(\frac{(\kalpha-1)r}{k} \right).
\end{align*}
Since the maximum degree of $\Lin(H)$ is at most $dk$, by~\cref{o:2tree} and taking a union bound over all possible $2$-trees of size $r$ that could contain fixed hyperedge $e$, we see that 
$$\P[\mcB_e] \le \frac{k}{2} (e^2 d^2 k^2)^{r-1} \cdot  \left(\frac12\right)^{(\kalpha-1)r} \exp\left(\frac{(\kalpha-1)r}{k} \right) \le \frac{1}{2e^2 d^2k} \left(\frac{2e^2d^2k^2}{2^{\kalpha}}\right)^r,$$
using that $\kalpha \le k$. Since $2^{\kalpha} \ge 4 e^2 d^2 k^2$, $\P[\mcB_e] \le \frac{1}{d 2^{r+1}}$ and thus plugging in $r$ and taking the sum gives $\P[\mcB_i] \le \frac{1}{n^2}$.
\end{proof}

\subsection{Looseness in the LLL regime}
We observe that the marking constructed in~\cref{s:boundcon} not only implies connectivity, but also implies \textit{looseness} in the sense of~\cite{ACH08} and~\cref{d:loose} for bounded degree CNFs.
Throughout this subsection, we fix a marking $\mcM \subset V$ satisfying the conditions of~\cref{lem:marking-robust}.

\begin{thm}[Looseness]
\label{t:lllloose}
There exist constants $\gamma \ge 0.1742$, $c > 0$, and $k_0>0$ such that the following holds for any integer $k \ge k_0$ and any $\zeta \in [0,1/2)$. 
Let $\Phi$ be a $(k,\zeta,d)$-CNF with $d \le \frac{c}{k^3} 2^{\gamma(1-\zeta) k}$. 
Suppose that $\pi$ is a distribution over solutions of $\Phi$ whose marginal distributions on marked variables $\pi_{\mcM}$ is $k$-locally uniform. 
Then whp a random assignment $\sigma$ chosen from $\pi$ is $O(dk^2 \log n)$-loose.
\end{thm}

\begin{proof}
By our choice of parameter regime and marking in~\cref{lem:marking-robust}, we have $2^{k_u} \ge 2 ed k^2 2^{k_m} \ge 4 e^2 d^2 k^2$.
Take $\sigma \sim \pi$ and some $v \in V$. 
We will show that with probability at least $1-O(1/n^2)$ over the choice of $\sigma$ there exists another assignment $\tau$ such that $\sigma(v) \neq \tau(v)$ and $\norm{\sigma - \tau}_1 = O(dk^2 \log n)$. 

Therefore, by a union bound, we see that with probability $1 - O(1/n)$, all variables $v \in V$ are $O(dk^2 \log n)$-loose with respect to $\sigma.$

Let $X = \sigma(\mcM \setminus \{v\})$ and let $\mcE_v^X$ be the connected component of $H_{\Phi^X}$ containing the variable $v$. 
We obtain a new assignment $\tau$ from $\sigma$ by updating variables in $\mcE_v^X$ so that $\tau(v) \neq \sigma(v)$. 
We note that it is always possible to make such a flip on $v$ conditioned on $X$ by \cref{l:vtxapproxuniform}. 
Hence, to show that $\tau$ satisfies our requirements it is enough to show that $|\mcE_v^X| = O(dk^2 \log n)$ with probability at least $1-O(1/n^2)$, which was already established in \cref{l:smallcomp}.
\end{proof}

\section{Random \texorpdfstring{$k$}{k}-CNFs}
\label{sec:random-CNF}

\subsection{Further preliminaries}
\begin{notn}
Throughout this section, unless otherwise specified, $\Phi$ denotes a uniformly random $k$-CNF on $n$ variables with $m = \alpha n$ clauses and $\Omega = \Omega_{\Phi}$ denotes the set of satisfying assignments of $\Phi$. Let $\Phi_k(m ,n)$ denote the set of $k$-CNFs with $n$ variables and $m = \alpha n$ clauses. We let $\Delta = k^4 \alpha$.
\end{notn}

\begin{defn} \label{d:goodbad}
A variable in $\Phi = (V, \mcC)$ is of \textit{high degree} if $\Phi$ contains at least $\Delta$ instances of literals involving $v$. Let $\HD(\Phi)$ denote the set of high degree variables of $\Phi.$ We define the \textit{bad variables} and \textit{bad clauses} of a formula $\Phi$ via an iterative process, similar to~\cite{GAL19}:
    \begin{itemize}
        \item $\mcV_0 \leftarrow \HD(\Phi)$ , $\mcC_0 \leftarrow$ clauses with at least $\zeta k$ high degree variables.
        \item Until $\mcV_i = \mcV_{i-1}$, repeat: increment $i$, let $\mcV_i \leftarrow \mcV_{i-1} \cup \text{vbl}(\mcC_{i-1})$ and $\mcC_i = \{c \in \mcC \mid \text{vbl}(c) \cap \mcV_i \ge \zeta k\}$
        \item $\mcC_{\bad}$ is the final set of bad clauses $\mcC_i$ and $\mcV_{\bad}$ is the final set of bad variables $\mcV_i$. Let $\mcV_{\good},\mcC_{\good}$ be the complements of the aforementioned sets.
    \end{itemize}
\end{defn}
Per the above definition, every good clause has less than $\zeta k$ bad variables, and any bad clause has \textit{only} bad variables.

\begin{defn} We associate several graphs and hypergraphs with $\Phi$.
\begin{itemize}
    \item Given $\Phi$, let $\mcH_{\Phi} = (V, \mcC)$ be the \textit{dependency (multi)hypergraph} where $V$ is the set of variables and $\mcC = \{ \text{var}(C) : C \in \Phi \}$ is the collection of clauses of $\Phi$ viewed as $k$-sets of variables.
    \item For a partial assignment $X \in \{0, 1\}^{\Lambda}$, let $\mcH_{\Phi}^X$ be the associated (not necessarily $k$-uniform) hypergraph to $\Phi^X$ and for variable $v \in V \setminus \Lambda$, let $\mcE_v^X$ denote the connected component of $\mcH_{\Phi}^X$ to which $v$ belongs.
    \item The \textit{clause dependency graph} $G_{\Phi}$ is a graph with vertex set the collection of clauses of $\Phi$, with an edge connecting clauses that share variables.
    \item The \textit{good clause dependency graph} $G_{\Phi, \good}$ has vertex set $\mcC_{\good}$ with edges between clauses that share a \textit{good} variable.
     \item The \textit{bad clause dependency graph} $G_{\Phi, \bad}$ is a graph with vertex set $\mcC_{\bad}$ with edges between bad clauses that  that share a (bad) variable.
     \item The \textit{variable dependency graph} $H_{\Phi}$ is a graph with vertex set the set of variables $V$, with an edge connecting variables that appear in some clause in $\Phi$ .
     \item The \textit{bad variable dependency graph} $H_{\Phi, \bad}$ connects variables that appear together in a bad clause. 
     \item A \textit{bad component} is a connected component of $H_{\Phi, \bad}$.
\end{itemize}
\end{defn}

We will leverage structural properties of bad components. In particular, we will need the following slight generalization of Lemma 48 in~\cite{GAL19} (that follows by a similar argument).

\begin{prop}
\label{l:randbadcomp}
With high probability over $\Phi$, every bad component $S$ has size at most $\frac{7}{\zeta} k \log n$.
\end{prop}
For completeness, we include a proof in~\cref{a:badcomps}.
In our algorithms, we will be particularly interested in the following induced formula associated to the low degree vertices of $\Phi$.

\begin{defn}
Let $\Phi_{\good} = (\mcV_{\good}, \mcC^-_{\good})$ be the \textit{induced good CNF} of $\Phi$ with variable set $\mcV_{\good}$, where for each $C \in \mcC_{\good}$, we associate a clause $C^- \in \mcC^-_{\good}$ by deleting any literals corresponding to bad variables in $C$. 
\end{defn}

\begin{rem}
Observe that each clause in $\Phi_{\good}$ has between $(1 - \zeta)k$ and $k$ distinct literals and that $\Phi_{\good}$ has at most $n$ variables. Further, the maximum variable degree of $\Phi_{\good}$ is $\Delta$.
\end{rem}

We observe the following result about marking:

\begin{prop}\label{l:randmark}
For some $\gamma \ge 0.1742$, if $\Delta  \lesssim \frac{2^{\gamma(1- \zeta)k}}{k^4}$, then with high probability over $\Phi$, there exists a set of marked variables $\mcM \subset \mcV_{\good}$ such that every clause in $\Phi_{\good}$ has at least $\kalpha := 2\gamma (1 - \zeta) k$ marked and $\kbeta := \gamma (1 - \zeta) k$ unmarked variables where we can partially assign all bad variables in a way to satisfy all clauses in $\mcC_{\bad}$.
\end{prop}
\begin{proof}
Observe that since $\Phi_{\good}$ is a CNF where every clause has between $(1-\zeta)k$ and $k$ literals with maximum degree $k\Delta$ by~\cref{l:alglll}, we can find a marking $\mcM \subset \mcV_{\good}$ such that every clause in $\Phi_{\good}$ has at least $\kalpha := 2\gamma (1 - \zeta) k$ marked and $\kbeta := \gamma (1 - \zeta) k$ unmarked variables. Further, since $\Phi$ is satisfying with high probability, we can take an arbitrary satisfying assignment $\sigma \in \Omega$ which must necessarily satisfy $\mcC_{\bad}$ and thus partial assignment $\sigma|_{\mcV_{\bad}}$ satisfies all of the clauses in $\mcC_{\bad}.$
\end{proof}

\subsection{\texorpdfstring{$O(\log n)$}{O(log n)} connectivity for random \texorpdfstring{$k$}{k}-CNF solutions}\label{s:randcon}

In this section we prove~\cref{t:rand-connected}, showing that two uniformly random solutions of a random $k$-CNF with bounded average degree $\alpha$, are $D$-connected for $D = O(\alpha k^6 \log n)$ with high probability; this implies the existence of a giant $O(\alpha k^6 \log n)$-connected component consisting of almost all solutions. 

Throughout this section, we suppose $\Phi \sim \Phi_k(m, n)$ is a random $k$-CNF with solution space $\Omega = \Omega_{\Phi}$ for $k$ sufficiently large and let $\mu = \mu_{\Phi}$ be the uniform distribution on $\Phi$. We employ the following further notation conventions in this subsection only.

\begin{notn}\label{n:rand}
\begin{itemize}
\item We fix arbitrarily small constant $\zeta \in (0, 1/2)$ such that $\gamma(1-\zeta) \ge 0.174$ for $\gamma$ as in~\cref{l:alglll}.
    \item We let $\kalpha = 2\gamma(1- \zeta)k, \kbeta = \gamma(1-\zeta)k$.
    \item We suppose $\Phi = (V, \mcC)$ with $n$ variables $V$ and $m = \alpha n$ clauses $\mcC$, for $\alpha \le  2^{k/6} \le \frac{1}{k^7} 2^{\gamma(1-\zeta)k}$. Let $d := \alpha k$.
    \item We choose $\Delta := k^4 \alpha \lesssim \frac{1}{k^4} 2^{\gamma(1-\zeta)k} $ 
    \item  We partition $V = \mcV_{\good} \sqcup \mcV_{\bad}$ and $\mcC = \mcC_{\good} \cup \mcC_{\bad}$ per~\cref{d:goodbad} with parameter $\zeta$ 
    \item Throughout the next two subsections, we fix some good marking $\mcM \subset \mcV_{\good}$ that satisfies the conditions of~\cref{l:randmark}.
\end{itemize}
\end{notn}

We consider the following algorithm for constructing a path of satisfying assignments between two random solutions so that the Hamming distance between adjacent assignments is $O(\Delta k^2 \log n)$. 

\begin{algorithm}[t]
\caption{Finding an $O(k^2 \Delta\log n)$-path between two solutions of random $k$-CNF $\Phi$}\label{alg:rCNF-path}
\KwIn{A $k$-CNF $\Phi \sim \Phi_k(m, n)$, high-degree threshold $\Delta$, a set $\mcM = \{v_1, \ldots, v_{\ell}\}$ of marked good variables, two solutions $\sigma,\sigma'$.} 

\medskip
Take a uniformly random solution $\psi$ of $\Phi_{\good}  = (\mcV_{\good}, \mcC^-_{\good})$\;
Run \cref{alg:find-path} to find an $O(\Delta k^2 \log n)$-path between $\sigma(\mcV_{\good})$ and $\psi$ in the solution space of $\Phi^X$ where $X = \sigma(\mcV_{\bad})$; we denote this path by
$\sigma(\mcV_{\good}) = \xi_0 \leftrightarrow \cdots \leftrightarrow \xi_\ell = \psi$
\Comment*[r]{Observation: $\psi$ is also a solution to $\Phi^X$ because $\mcC^X \subseteq \mcC^-_{\good}$}

Lift this path to the following $O(\Delta k^2 \log n)$-path between $\sigma$ and $\tau = \psi \cup \sigma(\mcV_{\bad})$ in the solution space of $\Phi$:
$\sigma = \zeta_0 \leftrightarrow \cdots \leftrightarrow \zeta_\ell = \tau$, where $\zeta_i = \xi_i \cup \sigma(\mcV_{\bad})$\;

Repeat the same steps for $\sigma'$ to obtain an $O(\Delta k^2 \log n)$-path between $\sigma'$ and $\tau' = \psi \cup \sigma'(\mcV_{\bad})$ in the solution space of $\Phi$, denoted by 
$\sigma' = \zeta'_0 \leftrightarrow \cdots \leftrightarrow \zeta'_{\ell'} = \tau'$\;

Construct a path between $\tau$ and $\tau'$ in $\Omega$ as follows:
\begin{itemize}
    \item Let $X = \tau(\mcV_{\good}) = \psi = \tau'(\mcV_{\good})$
    \item Suppose that $H_{\Phi}^X$ has connected components $\mcE_1, \ldots, \mcE_r$
    \item For each $i \in [r]$, construct $\tau_i \in \Omega$ where 
    $$\tau_i(v) = \begin{cases}
    \tau(v) & v \in \mcV_{\good} \\
    \tau(v) & v \in \bigcup_{j = i+1}^r \mcE_j \\
    \tau'(v) & v \in \bigcup_{j = 1}^i \mcE_j
    \end{cases}$$
    \item This yields path $\tau = \tau_0 \leftrightarrow \tau_1 \leftrightarrow \cdots \leftrightarrow \tau_r = \tau'$\;
\end{itemize}

\medskip

\KwOut{The path in $\Omega$
$$\sigma = \zeta_0 \leftrightarrow \cdots \leftrightarrow \zeta_\ell = \tau = \tau_0 \leftrightarrow \tau_1 \leftrightarrow \cdots \leftrightarrow \tau_r = \tau' = \zeta'_{\ell'} \leftrightarrow \cdots \leftrightarrow \zeta'_{0} = \sigma'$$
} 
\end{algorithm}

\cref{alg:rCNF-path} actually yields an explicit path of satisfying assignments of $\Phi$ (with high probability over $\Phi, \sigma, \sigma'$) that is computable in polynomial time, thereby algorithmically establishing connectivity.

\begin{lemma} \label{l:rand-well-def}
With high probability over $\Phi, \sigma, \sigma'$,~\cref{alg:rCNF-path} is well-defined in the following sense:
\begin{enumerate}
    \item We can find $\psi$ as in~\cref{alg:rCNF-path}.
    \item For all $i \in [\ell]$, there exists $\xi_i \in \Omega_{\Phi^X}$ as in~\cref{alg:rCNF-path} where the associated $\zeta_i = \xi_i \cup \sigma(\mcV_{\bad})$ is a satisfying assignment of $\Phi$.
    \item For all $i \in [\ell']$, there exists $\zeta_i' \in \Omega_{\Phi^X}$ as in~\cref{alg:rCNF-path}.
    \item For all $i \in [s]$, $\tau_i \in \Omega$ for $\tau_i$ defined as in~\cref{alg:rCNF-path}.
\end{enumerate}
\end{lemma}
\begin{proof}
We can find some $\psi$ as in~\cref{alg:rCNF-path} by applying the algorithmic Lov\'asz local lemma (\cref{p:lll}), since every clause in $\mcC_{\good}^-$ has at least $(1 - \zeta)k$ and at most $k$ variables, and the maximum degree of $\Phi_{\good}$ is $\Delta$.

We can apply~\cref{alg:find-path} with the marking $\mcM$ to $\Phi^X$ (where $X$ is as defined in~\cref{alg:rCNF-path}). Since every clause in $\mcC_{\good}^-$ has at least $(1 - \zeta)k$ clauses, by~\cref{t:robustconn}, $\xi_i \in \Omega_{\Phi^X}$. Further $\sigma(\mcV_{\bad})$ satisfies all clauses in $\mcC(\Phi) \setminus \Phi^X$ since $\sigma$ is a satisfying assignment and thus the assignments $\zeta_i \in \Omega$. We similarly have that $\zeta_i' \in \Omega$.

Observe that since $\tau, \tau'$ are satisfying assignments, the restrictions $\tau(\vbl(\mcE_i)), \tau'(\vbl(\mcE_i))$ must satisfy the clauses in $\mcE_i$ and the variables in different $\mcE_i$ are disjoint since each $\mcE_i$ is a connected component. Consequently, each $\tau_i \in \Omega$.
\end{proof}

\begin{lemma}
\label{lem:local-uni-RCNF}
Let $\Phi \sim \Phi_k(m, n)$ be a random $k$-CNF with $2^{(1 -\zeta)k} \ge 2 e \Delta k$. 
Let $X \in \{0, 1\}^{\mcV_{\bad}}$ be a partial assignment on bad variables that is extendable to a full satisfying assignment. 
Then $\mu_{\mcM}( \cdot | X)$ is $k$-locally uniform.
Moreover, $\mu_{\mcM}$ is $k$-locally uniform. 
\end{lemma}
\begin{proof} 
Fix some $X \in \{0, 1\}^{\mcV_{\bad}}$. We observe that $\mu(\cdot \mid X)$ is the uniform distribution on $\Phi^X$.
Since $\Phi^X$ is a $(k,\zeta,\Delta)$-CNF and $2^{(1 - \zeta) k} \ge 2e \Delta k$ so by \cref{l:vtxapproxuniform}, the marginal conditional distribution $\mu_{\mcM}( \cdot | X)$ is $k$-locally uniform.
Further, by the law of total probability,
\[
\mu_{\mcM}(\sigma(U) = \tau) 
= \sum_{X} \mu(X) \mu_{\mcM}(\sigma(U) = \tau | X),
\]
where the summation is over all feasible assignments $X \in \{0, 1\}^{\mcV_{\bad}}$ on bad variables.
Hence, the local uniformity of $\mu_{\mcM}$ follows from that of $\mu_{\mcM}( \cdot | X)$. 
\end{proof}

\begin{lemma}\label{l:rand-logn}
With high probability over the choices of $\Phi$ and two random solutions $\sigma, \sigma' \in \Omega$, the following properties hold (where assignments are defined as in~\cref{alg:rCNF-path}):
\begin{enumerate}
\item $\|\zeta_i - \zeta_{i-1}\|_1 = O(\Delta k^2 \log n)$ for all $i \in [\ell]$.
\item $\|\zeta_i' - \zeta_{i-1}'\|_1 = O(\Delta k^2 \log n)$ for all $i \in [\ell']$.
\item $\|\tau_i - \tau_{i-1}\|_1 = O(k \log n)$ for all $i \in [r]$.
\end{enumerate}
\end{lemma}
\begin{proof} 
We will apply~\cref{t:robustconn} with maximum degree $\Delta$, and parameters $\zeta, \gamma$. We observe that $\psi$ is a uniformly random solution of $\Phi_{\good}$. 
Note that every solution of $\Phi_{\good}$ is a satisfying assignment for $\Phi^X$ for any partial assignment $X \in \{0, 1\}^{\mcV_{\bad}}$. 
We observe that $\mu_{\Phi_{\good}}$ induces a $k$-locally uniform distribution on $\mcM$ by \cref{l:vtxapproxuniform}.

We consider a pair of uniformly random solutions $\sigma, \sigma' \in \Omega$, recalling that $\mu_{\mcM}(\cdot \mid X)$ for $X \in \{0, 1\}^{\mcV_{\bad}}$ and $\mu_{\mcM}$ are $k$-locally uniform by~\cref{lem:local-uni-RCNF}. Consequently $\sigma(\mcV_{\good})$ has $k$-locally uniform marginal on $\mcM$ for $(k,\zeta,\Delta)$-CNF $\Phi^X$ where $X = \sigma(\mcV_{\bad})$ and the same holds for $\sigma'$.

We apply~\cref{t:robustconn}. By first considering $(k, \zeta, \Delta)$-formula $\Phi^X$ for $X = \sigma(\mcV_{\bad})$ and assignments $\psi, \sigma(\mcV_{\good})$, with high probability, $\|\zeta_i - \zeta_{i-1}\|_1 = O(\Delta k^2 \log n)$ for all $i \in [\ell]$. We can analogously apply~\cref{t:robustconn} to  $(k, \zeta, \Delta)$-formula $\Phi^{X'}$ for $X' = \sigma'(\mcV_{\bad})$ and assignments $\psi, \sigma'(\mcV_{\good})$ and find that with high probability  $\|\zeta_i' - \zeta_{i-1}'\|_1 = O(\Delta k^2 \log n)$ for all $i \in [\ell']$

By~\cref{l:randbadcomp}, with high probability over $\Phi$, every bad component $S$ of $\Phi$ has size at most $\frac{7}{\zeta} k \log n$. This implies that $\|\tau_i - \tau_{i-1}\|_1 = O(k \log n)$ for all $i \in [r]$.
\end{proof}

\begin{proof}[Proof of~\cref{t:rand-connected}]
The desired result follows immediately from~\cref{alg:rCNF-path} enjoying the properties proved in~\cref{l:rand-well-def,l:rand-logn}.
\end{proof}

\subsection{Sampling random \texorpdfstring{$k$}{k}-CNF solutions via MCMC}
\label{subsec:sampling-rCNF}

In this section, we prove~\cref{t:sampling}

We will employ the following algorithm to sample a satisfying assignment to random $k$-CNF $\Phi \sim \Phi_k(m, n)$. 
The high-level idea is to run a heat-bath block dynamics on the set of marked variables. More precisely, we fix constant parameter $\theta \in (0,1)$. We design a Markov chain that will at every step, uniformly at random pick $\theta |\mcM|$ variables in $\mcM$ and update them conditioned on the values of the other assigned variables. 
It is easy to show that this dynamics is irreducible, aperiodic, and has stationary distribution $\mu_\mcM$. 

We work in a slightly different (sparser) density regime than the previous subsections. 
\begin{notn}\label{n:randsample}
\begin{itemize}
  \item Throughout this section, we suppose $\Phi \sim \Phi_k(m, n)$ is a random $k$-CNF with solution space $\Omega = \Omega_{\Phi}$ for $k$ sufficiently large. 
  \item Let $\mu = \mu_{\Phi}$ be the uniform distribution on $\Phi$.
    \item We suppose $\Phi = (V, \mcC)$ has $n$ variables $V$ and $m = \alpha n$ clauses $\mcC$, for $\alpha \le 2^{k/52}$. Let $d := \alpha k$.
    \item We let $\Delta = k^4 \alpha \le k^4 2^{k/52} $  and partition $V = \mcV_{\good} \sqcup \mcV_{\bad}$ and $\mcC = \mcC_{\good} \cup \mcC_{\bad}$ per~\cref{d:goodbad} with parameter $\zeta$ such that $\gamma(1-\zeta) \ge 0.174$ for $\gamma$ as in~\cref{l:alglll}.
    \item We fix a subset of marked variables $\mcM \subset \mcV_{\good}$ satisfying the conditions of~\cref{l:randmark} such that every $C \in \mcC_{\good}$ has at least $\kalpha = 2\gamma(1-\zeta)k$ marked and $\kbeta = \gamma(1 -\zeta)k$ unmarked variables.
    \item We define $\kgamma = \frac{4}{4(1-12\zeta) +5} \kbeta$ 
    \item We fix constant $\theta \in (0, 1)$ sufficiently small.
\end{itemize} 
\end{notn}

\begin{algorithm}[t]
\caption{The $\theta$-block dynamics $P_{\theta}$}  \label{a:randblockdynamics}
\KwIn{A $k$-CNF formula $\Phi = (V, \mcC)$, a set of marked (good) variables $\mcM \subset V$, and parameters $\zeta, \theta > 0$.} 

\medskip 
Fix a parameter $\eps > 0$ and let $T_{\max} \sim n^{C \Delta k^2}$ for some constant $C = C(\theta) > 0$\;

Initialize partial assignment $X_0$ by letting  $X_0(v) \overset{\text{R}}\leftarrow \{0, 1\}$ independently for each $v \in \mcM$\;

\For{$t = 1,\dots,T_{\max}$}{
	Choose a uniformly random subset $S \subset \mcM$ of size $|S| = \theta |\mcM|$\;
	For all  $v \not \in S$, set $X_t(v) \leftarrow X_{t-1}(v)$\;
	Set $X_t(S) \rand \mu_{S}(\cdot \mid X_{t-1}(\mcM \setminus S) )$\;
}

Extend $X_{T_{\max}}(\mcM)$ by choosing $X(V \setminus \mcM) \rand \mu_{V \setminus \mcM}(\cdot \mid X_{T_{\max}}(\mcM))$\;

\medskip

\KwOut{The assignment $X$ on all of $V$
} 
\end{algorithm}

In order to prove the correctness and the efficiency of \cref{a:randblockdynamics}, we need to show that with high probability over the choice of the random formula, the following facts are true: 
\begin{enumerate}
    \item We can efficiently find bad variables and marked variables, as already observed in previous sections;  (see~\cref{l:randmark,d:goodbad});
    \item The block dynamics, assuming flawless implementation, is rapidly mixing, i.e., the mixing time of the idealized block dynamics is polynomial in $n$;
    \item We can efficiently implement all (polynomially many) steps of the block dynamics with high probability over the choices of the dynamics;
    \item After the dynamics, we can extend a partial assignment on marked variables to a full assignment on all variables.
\end{enumerate}

Note that Step (1) has been done in previous sections.
To show rapid mixing of the block dynamics, we utilize a spectral independence approach. 

\begin{defn}
For any subset $\Lambda \subseteq \mcM$ and any pinning $\tau \in \{0,1\}^\Lambda$, the \textit{pairwise influence }from $u \in \mcM \setminus \Lambda$ to $v \in \mcM \setminus \Lambda$ is defined as
\[
\Psi^\tau(u,v) = \mu(X_v = 1 \mid X_u = 0, X_\Lambda = \tau) - \mu(X_v = 1 \mid X_u = 1, X_\Lambda = \tau) \quad \text{for $v \neq u$},
\]
and $\Psi^\tau(u,u) = 0$ for diagonal entries. 
The distribution $\mu_\mcM$ is called \textit{$\eta$-spectrally independent} if for any $\Lambda \subseteq \mcM$ and $\tau \in \{0,1\}^\Lambda$, the maximum eigenvalue of $\Psi^\tau$ is at most $\eta$. 
\end{defn}

\begin{prop}[\cite{AL20,ALO21,CLV21}]\label{p:blockmixes}
    If $\mu_\mcM$ is $\eta$-spectrally independent, then the (idealized) $\theta$-block dynamics has spectral gap at least $\theta^{O(\eta)}$ and mixing time at most $(1/\theta)^{O(\eta)} \log n$. 
\end{prop}

We then show that with a good marking, the marginal distribution on marked variables is $O(\log n)$-spectrally independent. 
\begin{lemma}\label{l:si}
Suppose that $\alpha \le \frac{1}{k^3} 2^{\frac{(1 -\zeta)(1-12\zeta)}{4(1-12\zeta)+5} \gamma k}$.
The marginal distribution $\mu_{\mcM}$ is $O(\Delta k^2 \log n)$-spectrally independent.
\end{lemma}

For steps (3) and (4), we show that we can efficiently implement the block dynamics and also extend to a full assignment.  

\begin{lemma}\label{l:appxblock}
Let $\alpha \le 2^{k/25}$ for sufficiently large $k$ for $\kalpha, \kbeta,\zeta$ chosen as in~\cref{n:randsample}.
With high probability, in each step and in the last step of~\cref{a:randblockdynamics}, every component has size $O(\Delta^2 k^3 \log n)$. 
\end{lemma}

We end this subsection with the proof of \cref{t:sampling}. 
The proof of~\cref{l:si} can be found in \cref{subsec:si}. 
The proof of~\cref{l:appxblock} can be found in \cref{subsec:implement}.

\begin{proof}[Proof of~\cref{t:sampling}]
Let $\Phi \sim \Phi_k(m, n)$ be a random $k$-CNF with $\alpha \le  2^{k/52}$ and $k$ sufficiently large.

We consider the choice of parameters in~\cref{n:randsample}. By assumption, $\alpha \le 2^{k/52}$. We have already observed that we can efficiently find bad and marked variables. 
Since $\alpha \le \frac{1}{k^3} 2^{\frac{\gamma(1 -\zeta)(1-12\zeta)}{4(1-12\zeta)+5} k}$, by~\cref{l:si}, we find that the marginal distribution $\mu_{\mcM}$ is $\eta$-spectrally independent for $\eta = O(\Delta k^2 \log n)$.

Consequently, applying~\cref{p:blockmixes}, we find that the idealized $\theta$-block dynamics mixes in at most $T_{\max} = O(n^{\Delta k^2})$ iterations. Subsequently applying~\cref{l:appxblock}, with high probability, we can implement all $T_{\max}$ steps of this dynamics such that in each step of~\cref{a:randblockdynamics}, every component has size $O(\log n).$ This implies that we can implement each iteration in time $2^{O(\log n)}$ (in fact, per~\cref{r:treeexcess}, we can compute marginals on components more efficiently using the low tree excess of the associated graphs/hypergraphs to a partially assigned random $k$-CNF). Further, by~\cref{l:appxblock}, we can efficiently extend our partial assignment on marked variables at the end of~\cref{a:randblockdynamics} to a full assignment, thereby enabling us to approximately sample from $\Omega_{\Phi}.$
\end{proof}

\subsubsection{Local uniformity and small components}

We begin by observing an analogue of~\cref{l:vtxapproxuniform} for random $k$-CNFs.
\begin{lemma}\label{l:randapproxunif}
Suppose that $2^{\kbeta} \ge 4e\Delta k$. Fix any partial assignment $X$ on $S \subset \mcM$. Then for any $v \in \mcV_{\good} \setminus S$ and $c \in \{0, 1\}$, we have 
$$\mu_v(c \mid X) \le \frac12 \exp\left( \frac{1}{k} \right)  < 1.$$
\end{lemma}
\begin{proof}
Fix some assignment $\Lambda$ of the variables in $\mcV_{\bad}$ that satisfies the clauses in $\mcC_{\bad}$. 
Let $\Phi^{\Lambda,X}$ be the CNF formula obtained by deleting all the clauses in $\Phi$ satisfied by $X \cup \Lambda$, noting that every clause $\Phi^{\Lambda, X}$ arises from a good clause in $\Phi$ by choice of $\Lambda$. Let $\mu^{\Lambda, X}$ be the uniform distribution on satisfying assignments of $\Phi^{\Lambda, X}$ and observe that $\mu_v^{\Lambda, X}(\cdot) = \mu_v(\cdot \mid \Lambda, x).$ Since $S \subset \mcM$, every clause left in $\Phi^{\Lambda,X}$ has between $\kbeta$ and $k$ variables, each of which belongs to at most $\Delta$ clauses. Since $2^{\kbeta} \ge 4e\Delta k$ by~\cref{p:lll}, for all $c \in \{0, 1\}$, we have that 
$$\mu_v^{\Lambda, X}(\cdot) = \mu_v(\cdot \mid \Lambda, X) \le \frac12 \exp\left(\frac{1}{k}\right).$$
Since the above holds for all $\Lambda$ and $\mu_v(\cdot \mid X)$ is a convex combination of the above $\mu_v(\cdot \mid \Lambda, X),$ the desired result follows.
\end{proof}

\begin{lemma}
Suppose $2^{\kbeta} \ge 4 e \Delta k$, let $\theta \le \frac13 \kalpha$, and take $k = k(\theta)$ sufficiently large.
For all $\theta \in (0, 1)$, $P_{\theta}$ as defined in~\cref{a:randblockdynamics} is an ergodic Markov chain supported on all of $\{0, 1\}^{\mcM}$ with unique stationary distribution $\mu_{\mcM}$ that it is reversible with respect to.
\end{lemma}

\begin{proof}
Similarly to~\cref{l:atklu}, if $2^{\kbeta} \ge 4 e \Delta k$, then for any $\Lambda \subset \mcM$ of size $|\Lambda| = \theta |\mcM|$, any update sequence $W \in \{0, 1\}^{\Lambda}$ assignment $Z \in \{0, 1\}^{\mcM \setminus \Lambda}$,
$$\mu(\Lambda = W \mid Z) \le \left(\frac12 \exp\left( \frac{1}{2k}\right) \right)^{|\Lambda|},$$
which in particular implies that for all choices of $W$, $\mu(\Lambda = W \mid Z) > 0$. This has the following consequences:
\begin{itemize}
    \item For any $X, Y \in \{0, 1\}^{\mcM}$, it is possible (with positive probability) to transform $X$ into $Y$ in $P_{\theta}$ with at most $\lceil \|X - Y\|_1 /(\theta|\mcM|) \rceil$ steps. Hence, $P_{\theta}$ is irreducible with support $\{0, 1\}^{\mcM}$.
    \item Since $P_{\theta}(X, X) > 0$ for any $X \in \{0, 1\}^{\mcM}$, $P_{\theta}$  is aperiodic. 
    \item To see that $P_{\theta}$ is reversible with respect to $\mu_{\mcM}$, consider any $X, Y \in \{0, 1\}^{\mcM}$ that differ only on a subset of  variables $\Lambda$ of size $|\Lambda| \le \theta |\mcM|$. Then,
    \begin{align*}
    \mu_{\mcM}(X) P_{\theta}(X, Y) &= \frac{1}{\binom{\mcM }{|\Lambda|}} \mu_{\mcM}(X) \mu_{\Lambda}( Y(\Lambda) \mid X(\mcM \setminus \Lambda)) \\
    &= \frac{1}{\binom{\mcM }{|\Lambda|}} \frac{\mu_{\mcM}(X)\mu_{\mcM}(Y)}{\mu_{\mcM \setminus \Lambda}(X(\mcM \setminus \Lambda)} \\
    &= \frac{1}{\binom{\mcM }{|\Lambda|}} \mu_{\mcM}(Y) \mu_{\Lambda}( X(\Lambda) \mid Y(\mcM \setminus \Lambda)) \\
    &= \mu_{\mcM}(Y) P_{\theta}(Y, X).
    \end{align*}
    Thus, $P_{\theta}$ has unique stationary distribution $\mu_{\mcM}$.
\end{itemize}
\end{proof}

We will show that the solution space of random $k$-CNF $\Phi$ shatters under a sufficiently balanced partial assignment. 
To do this, we will also leverage a generalization of a definition of~\cite{GAL19} and associated properties.

\begin{defn}\label{def:mcD}
For fixed positive integer $b \ge 2$, let $\mcD^{(b)}(\Phi)$ be the collection of subsets of clauses $T \subseteq \mcC$ such that the following pair of conditions hold:
\begin{itemize}
    \item $T \cap \mcC_{\good}$ is an independent set in $G_{\Phi,\good}$, i.e., for any pair of clauses $C_1, C_2 \in T$, one has $\vbl(C_1) \cap \vbl(C_2) \cap \mcV_{\good} = \emptyset$;
    \item The induced subgraph $G_{\Phi}^{\le b}[T]$ is connected, where   $G_{\Phi}^{\le b}$  is the graph where two vertices are adjacent if their distance is $\le b$ in $G_{\Phi}$. 
\end{itemize}
\end{defn}

We will first observe the following immediate generalization of Lemma 27 in~\cite{GAL19}.
\begin{lemma}\label{l:numts}
For fixed positive integer $b \ge 2$ and any positive integer $\ell \ge \log n$, with high probability over $\Phi$, every $C \in \mcC_{\good}$ has the property that the number of size $\ell$ subsets $T \in \mcD^{(b)}(\Phi)$ containing $C$ is at most $(18 k^2 \alpha)^{b \ell}$.
\end{lemma}

\begin{lemma}\label{c:dtreeexists-new}
Fix $\zeta \in (0, 1/2)$.
There exists constant $R = R(\zeta) > 0$ such that, with high probability over the randomness of $\Phi$, the following is true. 
For every subset of clauses $\mcE_0 \subseteq \mcC$ such that $|\mcE_0| \ge R \Delta k \log n$ and the induced subgraph $G_{\Phi}^{\le 2}[\mcE_0]$ is connected, 
and for every integer $\ell$ with $\ceil{R \log n} \le \ell \le \ceil{|\mcE_0|/(\Delta k)}$, 
we can find $T \subset \mcE_0$ that satisfies the following three conditions:
\begin{enumerate}[(a)]
    \item $T \in \mcD^{(4)}(\Phi)$;
    \item $|T| = \ell$;
    \item $|T \cap \mcC_{\good}| \ge (1 - 12\zeta) |T|$.
\end{enumerate}
\end{lemma}

To prove~\cref{c:dtreeexists-new}, we begin by recalling some properties of the graphs associated to $\Phi$, including a slight strengthening of Lemma 50 in~\cite{GAL19} that we prove in~\cref{a:badcomps}.

\begin{lemma}\label{l:fewbadclauses}
With high probability over $\Phi$, for any connected set of clauses $Y$ with $|\vbl(Y)| \ge \frac{8}{\zeta^3} k \log n$, we have that $|Y \cap \mcC_{\bad}|\le 3 \zeta |Y|$.
\end{lemma}

We also recall the following simple observation.
\begin{lemma}[Lemma 52~\cite{GAL19}]\label{l:conind}
Let $G$ be a connected graph and fix integer $b \ge 2$. For any connected induced subgraph $G'$ of $G^{\le b}$, there exists a connected induced subgraph of $G$ with size at most $b|V(G')|$ containing all vertices in $V(G')$.
\end{lemma}

\begin{lemma}[Lemma 2.3~\cite{CF14}]\label{l:expstrong}
For all sufficiently large $k$, whp over $\Phi$ for any $Y \subset \mcC$ with $|Y| \le n/k^2$, $|\vbl(Y)| \ge 0.9k|Y|$
\end{lemma}

\begin{proof}[Proof of~\cref{c:dtreeexists-new}]
Let $R > 0$ be sufficiently large which will be specified later.
Let $\mcE_{0,\good} = \mcE_0 \cap \mcC_{\good}$ be the set of good clauses in $\mcE_0$, and let $\mcE_{0,\bad} = \mcE_0 \setminus \mcE_{0,\good} = \mcE_0 \cap \mcC_{\bad}$ be the complement. 
Take $I \subseteq \mcE_{0,\good}$ to be an arbitrary \emph{maximal} independent set in $G_{\Phi,\good}[\mcE_{0,\good}]$. 
Note that every clause in $\mcE_{0,\good} \setminus I$ is adjacent to at least one clause in $I$ in the graph $G_{\Phi,\good}$ by the maximality of $I$. 
Since $G_{\Phi,\good}$ has maximum degree $(\Delta - 1)k$, 
we have that
\[
|I| \ge \frac{|\mcE_{0,\good}|}{(\Delta-1)k + 1} \ge \frac{|\mcE_{0,\good}|}{\Delta k}.
\]
Finally, we define $T = I \cup \mcE_{0,\bad} \subseteq \mcE_0$.
Our goal is to prove that $T$ (more precisely, a subset of $T$ of size exactly $\ell$) satisfies the three conditions in the lemma. 
\begin{enumerate}[(a)]
    \item We first show that $T \in \mcD^{(4)}(\Phi)$, 
    i.e., we need to verify the two conditions in \cref{def:mcD}. 
    The first condition is immediate from our construction since we take $I$ to be an independent set in $G_{\Phi,\good}$. 
For the second condition, 
suppose to the contrary $T$ was not connected in $G_{\Phi}^{\le 4}$. 
This would imply that we could partition $T = T' \sqcup T''$, where $d_{G_{\Phi}}(T', T'') \ge 5$. 
Since we know $\mcE_0$ is connected in $G^{\le 2}_\Phi$, let $C_0 \sim C_1 \sim \dots \sim C_\ell$
be a shortest path in $G^{\le 2}_\Phi$ such that $C_0 \in T'$, $C_\ell \in T''$, and $C_i \in \mcE_0 \setminus T$ for $1\le i < \ell$. 
Note that $\mcE_0 \setminus T$ contains only good clauses so every intermediate $C_i$ is a good clause. 
Consequently, each intermediate $C_i$ is adjacent to at least one clause in $I$ in $G_{\Phi,\good}$, and hence $C_i$ is adjacent to either $T'$ or $T''$ in the graph $G_\Phi$.
This immediately implies that $\ell \le 3$, as otherwise we can find a shorter path.  
If $\ell \le 2$, then 
\[
d_{G_{\Phi}}(T', T'') \le 2 \ell \le 4
\]
which is a contradiction.
It remains to consider the case $\ell = 3$. 
From the argument above we know that $C_1$ must be adjacent to $T'$ in $G_\Phi$, and $C_2$ is adjacent to $T''$. Thus,
\[
d_{G_{\Phi}}(T', T'') \le d_{G_{\Phi}}(T', C_1) + d_{G_{\Phi}}(C_1, C_2) + d_{G_{\Phi}}(C_2, T'') 
\le 1+2+1 = 4. 
\]
Again, this is a contradiction. Therefore, the second condition holds as well and $T \in \mcD^{(4)}(\Phi)$. 

 \item We next show that $T$ is relatively large:
 $$|T| = |I| + |\mcE_{0,\bad}| \ge \frac{|\mcE_{0,\good}|}{\Delta k} + |\mcE_{0,\bad}| \ge \frac{|\mcE_0|}{\Delta k} \ge R \log n.$$
 We can take a spanning tree of the subgraph $G_{\Phi}^{\le 4}[T]$ and remove leaf vertices to assume without loss of generality that $|T| = \ell$.
 
 \item Finally, we need to show that $|T \cap \mcC_{\good}| \ge (1 - 12\zeta)|T|$. 
 Since $T \in \mcD^{(4)}(\Phi)$, \cref{l:conind} implies that we can find a set $T \subset T' \subset V(G_{\Phi})$ with $|T'| \le 4|T|$ such that $G_{\Phi}[T']$ is connected. By \cref{l:expstrong},
 $$|\vbl(T')| \ge 0.9k|T'| \ge 0.9k|T| \ge 0.9R k \log n.$$
 By letting $R \ge \frac{80}{9 \zeta^3}$ be a sufficiently large constant, we can apply \cref{l:fewbadclauses} and observe that
 \begin{equation*}
     |T \cap \mcC_{\bad}| \le |T' \cap \mcC_{\bad}| \le 3\zeta |T'| \le 12 \zeta |T|.
 \end{equation*}
 Hence, we obtain $|T \cap \mcC_{\good}| \ge (1 - 12\zeta)|T|$. \qedhere
\end{enumerate}
\end{proof}

When $\mcE_0$ is connected, we can strengthen~\cref{c:dtreeexists-new}, by constructing $T \in \mcD^{(2)}(\Phi),$ with some more work, leveraging the following intermediate result.

\begin{lemma}\label{l:greenblue}
Let $G = (V,E)$ be a connected graph. 
Suppose that all vertices and edges are colored either green or blue. 
Let $V = V_{\mathrm{g}} \cup V_{\mathrm{b}}$ be the partition of green and blue vertices, and $E = E_{\mathrm{g}} \cup E_{\mathrm{b}}$ be for the edges, such that
\begin{itemize}
\item Every blue vertex is adjacent to only blue edges but no green edges;
\item Every green vertex can be adjacent to both green and blue edges, but only at most $D$ green edges. 
\end{itemize}
Then there exists a subset $T \subseteq V$ of vertices such that
\begin{enumerate}
\item $V_{\mathrm{b}} \subseteq T$;
\item $T \cap V_{\mathrm{g}}$ is an independent set in the subgraph $G[E_{\mathrm{g}}] = (V_{\mathrm{g}}, E_{\mathrm{g}})$ induced by green edges, and $|T \cap V_{\mathrm{g}}| \ge |V_{\mathrm{g}}|/(D+1)$;
\item $T$ is connected in $G^{\le 2}$.
\end{enumerate}
\end{lemma}

\begin{proof}
Observe that $E_g \subset E(G[V_{\mathrm{g}}])$, but $G[V_{\mathrm{g}}]$ may include some blue edges as well. We construct an independent set $I$ in the graph $G[E_{\mathrm{g}}]$ as follows. For each connected component $S$ in $G[V_{\mathrm{g}}]$, let $I_S \subset S$ be the independent set in $S$ chosen via the following procedure:
\begin{itemize}
    \item We consider the connected components of subgraph induced by $S$ and green edges, $G[E_{\mathrm{g}}, S]$. Let each connected component of $G[E_{\mathrm{g}}, S]$ initially be active and let $I_S = \emptyset$.
    \item Choose a first component $A_0$, find a maximal independent set and $2$-tree of $G[E_g, A_0]$, $I_{A_0}$, add $I_{A_0}$ to $I_S$, and mark component $A_0$ as inactive. This can be done by greedily adding vertices to the $2$-tree such that the newly added vertex is at distance exactly $2$ from the current set, till no vertex can be added (similar to the argument in~\cref{l:2treelin}).
    \item While some components are active, repeat the following
    \begin{itemize}
    \item Choose an active component $A \subset G[S] \cap E_{\mathrm{g}}$ that is adjacent in $G$ (via a blue edge $e = (a, b)$) to an inactive component $B$ (with $a \in A, b \in B$).
    \item Let $I_A$ be a maximal independent set of $G[A, E_{\mathrm{g}}]$, which we can choose to be a $2$-tree containing $a$, similarly to earlier.
    \item Add $I_A$ to $I_S$ and mark $A$ as inactive
    \end{itemize}
\end{itemize}
Since $S$ is a connected component, the above procedure must terminate. When the procedure terminates, $I_S$ is a maximal independent set of $G[S, E_{\mathrm{g}}]$ since each $I_A$ is a maximal independent set of $A$. 
Also, $I_S$ is connected in $G[S]^{\le 2}$ since each $I_A$ is a $2$-tree containing some vertex $a$ (except for $A_0$) that is adjacent via a blue edge to some vertex in the previous component $A'$, and this vertex must either be in $I_{A'}$ or be adjacent to some vertex in $I_{A'}$ by the maximality of $I_{A'}$.

We let $I = \bigcup_{S \text{ c.c.~in } G[V_{\mathrm{g}}]} I_S$, noting that $I$ is a maximal independent set in $G[E_g]$ and define $T = I \sqcup V_{\mathrm{b}}.$ We show that $T$ has the desired properties.
\begin{enumerate}[(1)]
    \item By choice $V_{\mathrm{b}} \subset T$.
    \item By construction $T \cap V_{\mathrm{g}} = I$, which is an independent set in $G[E_{\mathrm{g}}]$. Further, each green vertex can be adjacent to at most $D$ green edges and $I$ is a maximal independent set in $G[E_g]$, then
    $$|T \cap V_g| = |I| \ge \frac{|V_g|}{D + 1}.$$ 
    \item Suppose to the contrary that $T$ was not connected in $G^{\le 2}$. Then, we can partition $T = T' \sqcup T''$ where $d_G(T', T'') \ge 3$. Since $G$ is connected, we can find a shortest path $v_0 \rightarrow v_1 \rightarrow \cdots \rightarrow v_{\ell}$ in $G$ with $v_0 \in T', v_{\ell} \in T''$ and $v_i \in V \backslash T$ for each $i \in [\ell -1]$. Since $V_{\mathrm{b}} \subset T$, each intermediate $v_i \in V_{\mathrm{g}}$ for $i \in [\ell - 1]$. By maximality of $I$, every $v \in V_{\mathrm{g}}$ must either be in $I$ or adjacent to $I$ in $G[E_g]$. In particular, $v_1$ must be adjacent to a green vertex in $T'$ but not $T''$ since we chose a shortest path. Thus, we may assume that $v_0, v_{\ell}$ are green vertices, so $v_0, v_{\ell}$ are in the same connected component $S$ of $G[V_g]$. However, then $v_0, v_{\ell} \in I_S \subset I$ and are thus connected in $G[S]^{\le 2}$. This gives a contradiction. \qedhere
\end{enumerate}
\end{proof}

We can then conclude the following strengthening of~\cref{c:dtreeexists-new} that follows via a similar argument. We defer the proof to~\cref{a:missingsec3}.
\begin{lemma}\label{c:dtreeexists-conn}
Fix $\zeta \in (0, 1/2)$.
There exists constant $R = R(\zeta) > 0$ such that, with high probability over the randomness of $\Phi$, the following is true. 
For every connected (in $G_{\Phi}$) subset of clauses $\mcE_0 \subseteq \mcC$ such that $|\mcE_0| \ge R \Delta k \log n$, and for every integer $\ell$ with $\ceil{R \log n} \le \ell \le \ceil{|\mcE_0|/(\Delta k)}$, 
we can find $T \subset \mcE_0$ that satisfies the following three conditions:
\begin{enumerate}[(a)]
    \item $T \in \mcD^{(2)}(\Phi)$;
    \item $|T| = \ell$;
    \item $|T \cap \mcC_{\good}| \ge (1 - 6\zeta) |T|$.
\end{enumerate}
\end{lemma}

\subsubsection{Establishing spectral independence}
\label{subsec:si}

In this subsection we establish spectral independence and hence prove \cref{l:si}.

To show spectral independence, it suffices to show that under all \textit{pinnings} of subsets of marked variables, the expected Hamming distance is bounded under a suitable coupling.
Our coupling procedure is given in \cref{a:coupling}. 

\begin{algorithm}[ht!] 
\caption{Coupling $\mcC^{\tau}$ under pinning $\tau \subset \{0, 1\}^{\Lambda}$ for some $\Lambda \subseteq \mcM$} \label{a:coupling}
\KwIn{A $k$-CNF $\Phi$, a set of bad variables $\mcV_{\bad}$, a set of marked (good) variables $\mcM$, a partial assignment $\tau \in \{0, 1\}^{\Lambda}$ for $\Lambda \subset \mcM$, a parameter $\kgamma < \kbeta$, distinguished marked variable $v_0 \in \mcM \setminus \Lambda$.} 

\medskip
$X(\Lambda) \gets \tau$, $Y(\Lambda) \gets \tau$, $X(v_0) \gets 0$, $Y(v_0) \gets 1$ \Comment*{Initialization}
$\mcV_{\set} \gets \Lambda \cup \{v_0\}$ \Comment*{Set of (good) variables whose values have been revealed}
$\mcV_{\failed} \gets \{v_0\}$ \Comment*{Superset of uncoupled variables} 
$\mcE_{\failed} \gets \{\}$, $\mcE^\dagger_{\failed} \gets \{\}$, $\mcE^\ddagger_{\failed} \gets \{\}$ \Comment*{Three types of clauses causing failed variables} 

$\mcE_{\unsat} \gets \mcC(\Phi)$ \Comment*{Set of currently unsatisfied clauses}
\For{every $e \in \mcE$ satisfied by both $X(\mcV_{\set})$ and $Y(\mcV_{\set})$}{
	$\mcE_{\unsat} \gets \mcE_{\unsat} \setminus \{e\}$
	\Comment*{Remove clauses satisfied by the pinning $\tau$} 
	\label{line:remove-pinning}
}

\While{$\exists e \in \mcE_{\unsat}$ such that $\vbl(e) \cap \mcV_{\failed} \neq \emptyset$ and $(\vbl(e) \cap \mcV_{\good}) \setminus (\mcV_{\set} \cup \mcV_{\failed}) \neq \emptyset$ \label{line:whileloop}}{
	
	Pick an arbitrary good variable $u \in (\vbl(e) \cap \mcV_{\good}) \setminus (\mcV_{\set} \cup \mcV_{\failed})$:
	
	\qquad $r_u \rand [0, 1]$, $p_u^{X} \gets \mu(u = 1 \mid X(\mcV_{\set}))$, $p_u^{Y} \gets \mu(u = 1 \mid Y(\mcV_{\set}))$\;
	\label{line:opt-coupling}
	
	\qquad $X(u) \gets \ind(p_u^{X} \le r_u)$, $Y(u) \gets \ind(p_u^{Y} \le r_u)$\;

	\qquad $\mcV_{\set} \gets \mcV_{\set} \cup \{u\}$ \Comment*{Optimal coupling of $X(u)$ and $Y(u)$}

	\medskip
	\If{$X(u) \neq Y(u)$}{
		$\mcV_{\failed} \gets \mcV_{\failed} \cup \{u\}$, $\mcE_{\failed} \gets \mcE_{\failed} \cup \{e\}$ 
		\Comment*{$u$ is failed because of $e$} 
		\label{line:reason1}
	}

	\For{every $e \in \mcE$ satisfied by both $X(\mcV_{\set})$ and $Y(\mcV_{\set})$}{
		$\mcE_{\unsat} \gets \mcE_{\unsat} \setminus \{e\}$
		\Comment*{Remove clauses satisfied by the values of $u$} 
		\label{line:remove}
	}
	
	\For{$e \in \mcE_{\unsat}$ with $|\vbl(e) \cap \mcV_{\set} \setminus \Lambda| = \kgamma$}{
		$\mcV_{\failed} \gets \mcV_{\failed} \cup (\vbl(e) \setminus \mcV_{\set})$, $\mcE_{\failed} \gets \mcE_{\failed} \cup \{e\}$
		\Comment*{$e$ is failed because of the failure of local uniformity}
		\label{line:reason2} 
	}
	
	\For{$e \in \mcE_{\unsat}$ with $(\vbl(e) \cap \mcV_{\good}) \setminus (\mcV_{\set} \cup \mcV_{\failed}) = \emptyset$ and $(\vbl(e) \cap \mcV_{\bad}) \setminus \mcV_{\failed} \neq \emptyset$}{
		$\mcV_{\failed} \gets \mcV_{\failed} \cup (\vbl(e) \cap \mcV_{\bad})$, $\mcE^\dagger_{\failed} \gets \mcE^\dagger_{\failed} \cup \{e\}$
		\Comment*{$e$ is failed because of the failure of local uniformity} 
		\label{line:add-bad1}
	}
	
	\For{every connected component $F_{\bad}$ of $G_{\Phi, \bad}$ such that $\vbl(F_{\bad}) \cap \mcV_{\failed} \neq \emptyset$}{
		$\mcV_{\failed} \gets \mcV_{\failed} \cup \vbl(F_{\bad})$, $\mcE^\ddagger_{\failed} \gets \mcE^\ddagger_{\failed} \cup F_{\bad}$
		\Comment*{Bad components are assumed to be failed} 
		\label{line:add-bad2}
	}
}

$\mcV_{\coupled} \gets V \setminus \mcV_{\failed}$ 
\Comment*{Set of coupled vertices}

Extend $X(\mcV_{\set})$ and $Y(\mcV_{\set})$ to the same assignment on $\mcV_{\coupled} \setminus \mcV_{\set}$, and then
to the full assignment $X$ and $Y$ on $\mcV_{\failed} \setminus \mcV_{\set}$ with an arbitrary coupling.
\label{line:final-couple}

\medskip

\KwOut{$(X,Y)$}
\end{algorithm}

The coupling procedure, roughly speaking, operates by revealing the values of some of \emph{good} variables in a pair of assignments of $\Phi$ in a specifically chosen order. 
Whenever $X$ and $Y$ do not couple at a variable $u$, we put $u$ into $\mcV_{\failed}$ which is a superset of all variables that are or potentially will be uncoupled, and then move on and try to couple (good) neighbors of $u$.  

When coupling $u$, we always use the optimal coupling for the marginal distributions at $u$ conditioned on previously determined values. 
The local uniformity ensures that the \textit{uncouple probability}, i.e. the chance that $X$ and $Y$ assign $u$ differently, is tiny.
Whenever we set the values of a variable in both $X$ and $Y$, we always try to simplify the formula by removing all clauses that are satisfied in both $X$ and $Y$ using the currently determined variables from $\mcV_{\set}$. 
The coupling process stops when there are no unset variables adjacent to failed vertices.

Ideally, the process should stop soon since the uncouple probability can be made much smaller than the neighborhood growth or say the maximum degree of variables. 
There are two special events, however, that can harm the coupling procedure.
The first is that we have determined values of many variables in a clause $e$ but it is still not satisfied yet. Then there are no nice guarantees on the coupling probabilities of the remaining variables in $e$ since local uniformity will fail.
In this case we simply view all the remaining vertices as failed and add them into $\mcV_{\failed}$. 
Note that we have not set their values yet so they are not in $\mcV_{\set}$. 
The second bad case is that because of the first case, some bad variables might be included in $\mcV_{\failed}$, i.e., they potentially are discrepancies in the coupling. Since we have no good control on either the marginal probabilities or the degrees of the bad variables (no local uniformity), if this happens, we will simply regard the entire bad connected component (comprising only bad variables) as (potentially) uncoupled and add them to $\mcV_{\failed}$. 
Again we do not set their values yet so they are not contained in $\mcV_{\set}$.

The following fact is immediate from our coupling procedure.

\begin{lemma}
The following properties hold for the coupling $\mcC^{\tau}$ in~\cref{a:coupling}:
\begin{enumerate}
    \item Every clause $e \in \mcC(\Phi)$ satisfies at least one of the following:
    \begin{enumerate}
    \item $\vbl(e) \subseteq \mcV_{\set} \cup \mcV_{\coupled}$;
    \item $\vbl(e) \subseteq \mcV_{\set} \cup \mcV_{\failed}$;
    \item $e$ is satisfied by both $X(\mcV_{\set})$ and $Y(\mcV_{\set})$. 
    \end{enumerate}
    In particular, each clause in either $\Phi^{X(\mcV_{\set})}$ or $\Phi^{Y(\mcV_{\set})}$ has either all variables in $\mcV_{\coupled}$ or all in $\mcV_{\failed}$. 
    
    \item The coupling terminates eventually and returns a pair $(X,Y)$ such that $X$ is distributed as $\mu(\cdot \mid X(v_0) = 0, X(\Lambda) = \tau)$ and $Y$ is distributed as $\mu(\cdot \mid Y(v_0) = 1, Y(\Lambda) = \tau)$; furthermore, 
    \[
    X(\mcV_{\coupled}) = Y(\mcV_{\coupled}).
    \]
\end{enumerate}
\end{lemma}

\begin{proof}
\begin{enumerate}
\item Let $e$ be an arbitrary clause of $\Phi$ and assume that $e$ is not satisfied by both $X(\mcV_{\set})$ and $Y(\mcV_{\set})$. 
In particular, it is never removed in Lines~\ref{line:remove-pinning} and \ref{line:remove} in \cref{a:coupling}. 
If $e$ is a bad clause, then either $e$ never has a failed variable or all variables in $e$ become failed at some point by Line~\ref{line:add-bad2}. 
Assume that $e$ is a good clause. 
If $e$ does not contain any failed variables, then $\vbl(e) \subseteq \mcV_{\set} \cup \mcV_{\coupled}$.
Otherwise, it will enter the while loop in Line~\ref{line:whileloop} and the algorithm will try to set the values of variables in $e$ in both copies $X$ and $Y$, until when:
\begin{enumerate}
\item We have settled $\kgamma$ unpinned good variables, i.e., $|\vbl(e) \cap \mcV_{\set} \setminus \Lambda| = \kgamma$. Then all the remaining variables in $e$ will be added to $\mcV_{\failed}$ by Line~\ref{line:reason2}. 
\item All good variables have been either set or failed while some bad variables are undermined (not failed), i.e., $(\vbl(e) \cap \mcV_{\good}) \setminus (\mcV_{\set} \cup \mcV_{\failed}) = \emptyset$ and $(\vbl(e) \cap \mcV_{\bad}) \setminus \mcV_{\failed} \neq \emptyset$, then the remaining variables will be included to $\mcV_{\failed}$ by Line~\ref{line:add-bad1}.
\end{enumerate}
In both cases, eventually it holds $\vbl(e) \subseteq \mcV_{\set} \cup \mcV_{\failed}$ at the end of the algorithm, as claimed. 
As a corollary, each clause $e$ is either removed from the simplified formula $\Phi^{X(\mcV_{\set})}$ or contains only variables in $\mcV_{\coupled}$ or contains only variables in $\mcV_{\failed}$. 
This means that the hypergraph of $\Phi^{X(\mcV_{\set})}$ can be partitioned into two disconnected set $\mcV_{\coupled} \setminus \mcV_{\set}$ and $\mcV_{\failed} \setminus \mcV_{\set}$. The same holds for $\Phi^{Y(\mcV_{\set})}$ as well.

\item From (1) we know that $\Phi^{X(\mcV_{\set})}$ and $\Phi^{Y(\mcV_{\set})}$ induce the same simplified formula on $\mcV_{\coupled} \setminus \mcV_{\set}$ so we can couple them perfectly. 
Hence, by Line~\ref{line:final-couple} the output $(X,Y)$ is a coupling from the two target distributions. 
Moreover, all variables in $\mcV_{\coupled} \setminus \mcV_{\set}$ and in $\mcV_{\set} \setminus \mcV_{\failed}$ are the same in both $X$ and $Y$, and hence we have $X(\mcV_{\coupled}) = Y(\mcV_{\coupled})$. \qedhere
\end{enumerate}
\end{proof}

The discrepancies of $X$ and $Y$ all belong to the set $\mcV_{\failed}$.
To analyze the size of $\mcV_{\failed}$, we will instead consider $\mcE_{\failed}$, the set of failed clauses that cause failed vertices.
The following lemma is helpful for understanding the relation between $\mcV_{\failed}$ and $\mcE_{\failed}$. 
\begin{lemma}
\label{lem:C-E-failed}
The following properties hold for the coupling $\mcC^{\tau}$ in~\cref{a:coupling}:
\begin{enumerate}
    \item Every failed variable $v \in \mcV_{\failed}$ is contained in at least one failed clause from $\mcE_{\failed} \cup \mcE^\dagger_{\failed} \cup \mcE^\ddagger_{\failed}$,
    and in particular every failed good variable $v \in \mcV_{\failed} \cap \mcV_{\good}$ is contained in at least one failed clause from $\mcE_{\failed}$;
    \item The set $\mcE_{\failed} \cup \mcE^\dagger_{\failed} \cup \mcE^\ddagger_{\failed}$ of failed clauses is connected in $G_\Phi$;
    \item The set $\mcE_{\failed} \cup \mcE^\ddagger_{\failed}$ of failed clauses is connected in $G_{\Phi}^{\le 2}$;
\end{enumerate}
\end{lemma}

\begin{proof}
\begin{enumerate}
\item It is clear from the algorithm that whenever we add a failed variable to $\mcV_{\failed}$ we also add the corresponding failed clause causing it to one of $\{ \mcE_{\failed}, \mcE^\dagger_{\failed}, \mcE^\ddagger_{\failed}\}$, as in Lines~\ref{line:reason1}, \ref{line:reason2}, \ref{line:add-bad1}, and \ref{line:add-bad2}. 
Furthermore, when a good variable becomes failed we add the corresponding failed clause to $\mcE_{\failed}$, as in Lines~\ref{line:reason1} and \ref{line:reason2}.

\item Note that every time a new failed variable and failed clause is introduced, it is because some variables in this clause has already been failed and so the new failed clause is adjacent to a previous failed clause.

\item To see this, every time we introduce a failed clause in $\mcE^\dagger_{\failed}$, there must be some good variable $v$ in this clause that belongs to $(\mcV_{\failed} \setminus \mcV_{\set}) \cap \mcV_{\good}$, and hence $v$ is contained in some clause from $\mcE_{\failed}$. 
This implies that the set $\mcE_{\failed} \cup \mcE^\ddagger_{\failed}$ is connected in $G_{\Phi}^{\le 2}$. 
\qedhere
\end{enumerate}
\end{proof}

Our goal then is to bound the expected size of $\mcE_{\failed}$, which is given by the following lemma. 

\begin{lemma}
\label{lem:failed-coupling-component}
Suppose that $\alpha \le \frac{1}{k^3} 2^{\frac{1-12\zeta}{4} \kgamma}$ with $\kgamma \ge 1$ such that $2^{\kbeta - \kgamma} \ge 2 e \Delta s$ for $s \ge k$ satisfying $ s \ge k^{10} \alpha^{4/(1-12\zeta)} $. 
Then, with high probability over $\Phi$, for all $\tau \in \{0, 1\}^{\Lambda}$ for all $\Lambda \subset \mcM$, the coupling $\mcC^{\tau}$ defined in~\cref{a:coupling} satisfies $\EE[|\mcE_{\failed}|] = O(\Delta k \log n)$. 
\end{lemma}

We now show how to deduce \cref{l:si} from \cref{lem:failed-coupling-component}. 

\begin{proof}[Proof of~\cref{l:si}]
Let $\kgamma = \frac{4}{4(1-12\zeta) + 5} \kbeta$ for $\kalpha, \kbeta, \zeta$ given as in the beginning of \cref{subsec:sampling-rCNF}. Pick $s = \alpha^{4/(1-12\zeta)} k^{10}$ and observe that $\alpha \le \frac{1}{k^3} 2^{\frac{1 - 12\zeta}{4} \kgamma}$ and $2^{\kbeta - \kgamma} \ge 2 e \Delta s$ by assumption on $\alpha.$
Let $\Lambda \subset \mcM$ and fix arbitrary pinning $\tau \in \{0, 1\}^{\Lambda}.$
Let $u \in \mcM \setminus \Lambda$ be a marked variable. 
We will bound the sum of absolute influences from $u$ to all other marked variables, which provides an upper bound on the spectral independence constant.
Let $\mcC^\tau$ be the coupling from~\cref{a:coupling}. 
Then notice that
\begin{align*}
&\sum_{v \in \mcM \setminus \{u\}} |\Psi(u,v)| \\
={}& \sum_{v \in \mcM \setminus \{u\}} |\mu_\mcM(Y(v) = 1 \mid Y(u) = 1, Y(\Lambda) = \tau) - \mu_\mcM(X(v) = 1 \mid X(u) = 0, X(\Lambda) = \tau)| \\
\le{}& \sum_{v \in \mcM \setminus \{u\}} \Pr_{\mcC^{\tau}}(X(v) \neq Y(v)) \\
={}& \E_{\mcC^{\tau}} [d_{\mathrm{Ham}}(X(\mcM),Y(\mcM))] - 1.
\end{align*}
Our choice of parameters above satisfies~\cref{lem:failed-coupling-component} and thus we obtain that
\[
\E_{\mcC^{\tau}} [d_{\mathrm{Ham}}(X(\mcM),Y(\mcM))]
\le \EE[ k |\mcE_{\failed}|]
= O(\Delta k^2 \log n),
\]
as wanted.
\end{proof}

We leverage local uniformity to analyze the above coupling.

\begin{lemma}\label{l:plowhigh}
Suppose that $2^{\kbeta - \kgamma} \ge 2 e \Delta s$ for some $s \ge k$.
In~\cref{a:coupling}, if $u \in \mcV_{\good} \setminus \Lambda$, then 
$$
\frac12 - \frac{1}{s} 
\le 1 - \frac12\exp\left(\frac{1}{s}\right) 
\le p_u^X, p_u^Y 
\le \frac12\exp\left(\frac{1}{s}\right) 
\le \frac12 + \frac{1}{s}.
$$
In particular, if $r_u \le \frac12 - \frac{1}{s}$ or $r_u \ge \frac12 + \frac{1}{s}$, then $X(u) = Y(u)$.
\end{lemma}

\begin{proof}
By the initialization step of~\cref{a:coupling}, all pinned variables have the same assigned value. 
If $u \not \in \Lambda$, we show that $1 - \frac12\exp\left(\frac{1}{s}\right) \le p_u^X \le \frac12\exp\left(\frac{1}{s}\right)$, concluding the analogous result holds for $p_u^Y$. By Line~\ref{line:opt-coupling}, the distribution $\mu(u = \cdot \mid X, \tau)$ is the distribution $\mu$ conditioned on pinning the variables in $\mcV_{\set}$ where $\mcV_{\set}$ is the set of variables whose values have been revealed up to then (rather than till the end). For each clause $C \in \mcC,$ we observe that one of the following holds:
\begin{itemize}
    \item $C \in \mcC_{\bad}$;
    \item $C$ is satisfied by $X(\mcV_{\set})$;
    \item $|\vbl(c) \cap (\mcV_{\set} \setminus \Lambda)| < \kgamma$.
\end{itemize}
Since each clause in $\Phi$ has at least $\kbeta$ unmarked good variables, each clause remaining in $\Phi^{X(\mcV_{\set})}$ must have at least $\kbeta - \kgamma$ unmarked, good variables unassigned by $X(\mcV_{\set})$. Since $2^{\kbeta - \kgamma} \ge 2 e d s$, by~\cref{p:lll}, $\max\{p_u^X, 1- p_u^X \} \le \frac12 \exp(\frac{1}{s})$, and the results follows additionally for $Y$.
\end{proof}

\begin{proof}[Proof of \cref{lem:failed-coupling-component}]
Suppose that~\cref{a:coupling} terminates with partial assignment $X \in \{0, 1\}^{\mcV_{\set}}$. 
Note that $\mcE_{\failed} \cup \mcE^\ddagger_{\failed}$ is a connected set in $G^{\le 2}_\Phi$ by \cref{lem:C-E-failed}. We will upper bound $\P(|\mcE_{\failed}| \ge R \Delta k \log n)$. 
If $|\mcE_{\failed}| \ge R \Delta k \log n$, then $|\mcE_{\failed} \cup \mcE^\ddagger_{\failed}| \ge R \Delta k \log n$ and by~\cref{c:dtreeexists-new} we can find $T \subset \mcE_{\failed} \cup \mcE^\ddagger_{\failed}$ satisfying the conditions of~\cref{c:dtreeexists-new}.

We further restrict our attention to $T$ since
\begin{align*}
\P(|\mcE_{\failed}| \ge R \Delta k \log n) 
\le \sum_{T \in \mcD(\Phi)} \P(T\cap \mcC_{\good} \subseteq \mcE_{\failed}). 
\end{align*}
For each $T$, let $T_{\good} \subset T \cap \mcC_{\good}$ be the good clauses in $T$.
By~\cref{c:dtreeexists-new},  if $R$ is sufficiently large, then $|T| \ge \mcE_{\failed}|/(k\Delta) \ge R \log n$ and $|T_{\good}| \ge (1 - 12\zeta) |T|$.

Per~\cref{a:coupling,l:plowhigh}, there are two ways that a good clause $C$ can be part of $\mcE_{\failed}$:
\begin{enumerate}
    \item There exists $u \in (\vbl(C) \cap \mcV_{\set}) \setminus \{v_0\}$ such that $\frac12 - \frac{1}{s} \le r_u \le \frac12 + \frac{1}{s}$;
    \item $|\vbl(C) \cap (\mcV_{\set} \setminus \Lambda)| = \kgamma$ and $C$ is not satisfied by both $X(\mcV_{\agree})$ and $Y(\mcV_{\agree})$.
\end{enumerate}
Consequently, for every good clause $C \in \mcE_{\failed} \cap \mcC_{\good}$, either some variable in that good clause satisfies (1) above, or there are $\kgamma$ variables in $\vbl(C) \setminus \Lambda$ in $\mcV_{\set}$.
In particular, consider the random number $r_u$ selected in Line~\ref{line:opt-coupling} of \cref{a:coupling}. 
For a given $T_{\good}$, we can think of the coupling $\mcC^\tau$ as follows. 
Every good clause $C$ in $T_{\good}$ prepares $\kgamma$ independent random numbers, and whenever the algorithm would like to use a random number $r_u$ in Line~\ref{line:opt-coupling} for a variable $u$ in some (unique) good clause $C \in T_{\good}$, it takes the next unused random number $r$ that $C$ owns and let $r_u = r$ or $r_u = 1-r$ so that $X(u) = \varphi_C(u)$ if and only if $r \le \mu(u = \varphi_C(u) \mid X(\mcV_{\set}))$, where $\varphi_C$ is the assignment that $C$ forbids.
If (1) happens then one of the random numbers that $C$ owns satisfies $\frac12 - \frac{1}{s} \le r \le \frac12 + \frac{1}{s}$.
If (2) happens then all the $\kgamma$ random numbers of $C$ satisfy $r \le \frac12 \exp\left(\frac{1}{s}\right)$.

Since all clauses in $T_{\good}$ have disjoint sets of good variables and all the random numbers of clauses in $T_{\good}$ are independent of each other, analyzing $T_{\good}$ yields that
\begin{align*}
\P(T_{\good} \subseteq \mcE_{\failed}) 
&\le \left(\frac{2k}{s} + \left(\frac12 \exp\left(\frac{1}{s} \right)\right)^{\kgamma} \right)^{|T_{\good}|} \\
&\le \left(\frac{2k}{s} + \left(\frac12 \exp\left(\frac{1}{s} \right)\right)^{\kgamma} \right)^{(1 - 12\zeta)|T|}.
\end{align*}
Therefore, it follows from \cref{l:numts} that
\begin{align*}
\P(|\mcE_{\failed}| \ge R \Delta k \log n)
&\le (18 k^2 \alpha)^{4|T|}  \left(\frac{2k}{s} + \left(\frac12 \exp\left(\frac{1}{s} \right)\right)^{\kgamma} \right)^{(1 - 12\zeta)|T|} \\
&\le (18^4 k^8 \alpha^4)^{|T|}  \left(\frac{(2k)^{1 - 12\zeta}}{s^{1 - 12\zeta}} + \frac{1}{2^{(1 - 12\zeta)\kgamma}} \exp\left(\frac{(1-12\zeta)\kgamma}{s} \right) \right)^{|T|} \\
&\le  \left(\frac{18^4 (2k)^{1 - 12\zeta}k^8 \alpha^4}{s^{1 - 12\zeta}} + \frac{18^4 k^8 \alpha^4}{2^{(1 - 12\zeta)\kgamma}} \exp\left(\frac{(1-12\zeta)\kgamma}{s} \right) \right)^{|T|} \\
&= o\left(\frac{1}{n^R}\right),
\end{align*}
where the final inequality follows by $ \log n \lesssim |T| \ll n$, $s \ge \alpha^{4/(1-12\zeta)} k^{10} \gg \alpha^{4/(1-12\zeta)} k^{9/(1-12\zeta)}$, and $\alpha \le \frac{1}{k^3} 2^{\frac16 \kgamma} \ll \frac{1}{k^2} 2^{\frac{1-12\zeta}{4} \kgamma}$.

Since $|\mcE_{\failed}| \le m \le d n$, the result immediately follows, as 
$$\EE[|\mcE_{\failed}|] \le R \Delta k \log n + dn \P(|\mcE_{\failed}| > R\Delta k \log n) = O(\Delta k\log n).$$
This proves the lemma. 
\end{proof}

\subsubsection{Implementation of the block dynamics}
\label{subsec:implement}

In this subsection we give our proof of \cref{l:appxblock}.

We will leverage the \textit{local uniformity} of the distribution of satisfying assignments, a consequence of applying the Lov\'asz local lemma per~\cref{p:lll} to subformulae of $\Phi_{\good}.$

\begin{lemma}\label{l:atklu}
Suppose $2^{\kbeta} \ge 2 edk$. 
For every $t \ge 0$, the marginal distribution $X_t(\mcM)$ on marked variables (as defined in~\cref{a:randblockdynamics}) is $k$-locally uniform.
\end{lemma}
\begin{proof}
Observe that $X_0$ is initialized to be from the uniform distribution over $\{0,1\}^\mcM$ and hence the local uniformity of $X_0$ is trivial. Suppose $t \ge 1$ now, let $U \subset \mcM$ and fix a partial assignment $\tau \in \{0,1\}^U$ on $U$.
For each step of the block dynamics in \cref{a:randblockdynamics}, the algorithm picks a random block $S \subseteq \mcM$ and makes heat-bath update. 
We may assume that all marked variables are ordered and denoted by $\mcM = \{v_1,\dots,v_{|\mcM|}\}$.
In each step, when the algorithm tries to update a block $S = \{v_{i_1}, \dots, v_{i_\ell}\}$ where $\ell = \theta |\mcM|$ and $i_1 < \dots < i_\ell$, 
it makes updates sequentially under this ordering of variables, in the sense that for $j = 1,\dots, \ell$ the algorithm picks $X(v_{i_j})$ from the conditional distribution $\mu(X(v_{i_j}) = \cdot \mid X(\mcM \setminus S), X(\{v_{i_1}, \dots, v_{i_{j-1}}\}))$, i.e., conditioned on the values of $\mcM \setminus S$ and also $v_{i_1}, \dots, v_{i_{j-1}}$ that have already been picked. 
Denote the (random) sequence of blocks selected by the algorithm up to time $t$ by $S_1,\dots,S_t$ and fix an arbitrary one. 
For any variable $u \in U$, let $t_u$ be the last time that $u$ is updated, i.e., $t_u$ is the largest $t' \le t$ such that $u \in S_{t'}$ if such $t'$ exists and otherwise $t_u = 0$. 
We may assume that the algorithm samples an independent uniformly random number $r_u \in [0,1]$ at time $t_u$ and sets $X_{t_u}(u) = \tau(u)$ if $t_u \ge 1$ and $r_u \le \mu(X_{t_u}(u) = \tau(u) \mid X_{t_u}(\mcM \setminus S), X_{t_u}(S_{t_u, <u}))$ where $S_{t_u, <u} \subseteq S_{t_u}$ contains all variables less than $u$ (so they are updated before $u$), or if $t_u=0$ and $r_u \le 1/2$. 
By \cref{l:randapproxunif}, we deduce that $X_t(u) = X_{t_u}(u) = \tau(u)$ only if
\[
r_u \le \frac12 \exp\left(\frac{1}{k}\right).
\]
Hence, we obtain that
\begin{align*}
\P( X_t(U) = \tau)
&\le \max_{S_1,\dots,S_t} \P( X_t(U) = \tau \mid S_1,\dots,S_t) \\
&\le \max_{S_1,\dots,S_t} \P\left( \forall u \in U, \, r_u \le \frac12 \exp\left(\frac{1}{k}\right) \,\bigg\vert\, S_1,\dots,S_t \right) \\
&= \left(\frac12 \exp\left(\frac{1}{k}\right) \right)^{|U|},
\end{align*}
where the last equality is because all random numbers $r_u$'s are independent. This establishes the local uniformity.
\end{proof}

We now present our proof of \cref{l:appxblock}.

\begin{proof}[Proof of~\cref{l:appxblock}]
Suppose that at time $t$, we compute $X_t$ by resampling the vertices in $S \subset \mcM$ for some $S$ of size $|S| = \theta|\mcM|$. 
Let $X = X_{t-1}(\mcM \setminus S)$.
Let $\mcB_t$ be the bad event that $G_{\Phi^{X}}$ has some component of size at least $ R \Delta^2 k^3 \log n$ for sufficiently large constant $R$.
Suppose that bad event $\mcB_t$ occurs. Then, there is some connected component $\mcE^{X}$ with $|\mcE^{X} | \ge R \Delta^2 k^3 \log n$ where we view $\mcE^{X}$ as a collection of connected clauses without removing variables from $\mcM \setminus S$.
For sufficiently large $R$, we can find  $T \subset \mcE^{X}$ satisfying the conditions of~\cref{c:dtreeexists-conn}; in particular we let $|T| = \ceil{R \Delta k^2 \log n}$. 
Let $T_{\good} = T \cap \mcC_{\good}$ be the set of good clauses in $T$. 
Note that $T_{\good} \neq \emptyset$ for $R$ large enough since every component of bad clauses in the line graph $G_{\Phi}$ has size $O(\log n)$ by \cref{l:randbadcomp,l:expstrong}. 
Hence, we obtain that
\[
\P(\mcB_t) 
\le \sum_{T\text{ satisfying~\cref{c:dtreeexists-conn}}} \P(\text{all clauses in }T_{\good}\text{ unsatisfied by }X). 
\] 
Now fix a $T \in \mcD^{(2)}(\Phi)$ satisfying \cref{c:dtreeexists-conn}. 
For $\xi>0$, we say that the subset $S \subset |\mcM|$ of size $\theta |\mcM|$ is \textit{$(T,\xi)$-bad} if $|\vbl(T_{\good}) \cap S| \ge \xi \theta k |T|$ for some $\xi \ge 1$. 
Then we have
\begin{multline}\label{eq:first-second}
\P(\text{all clauses in }T_{\good}\text{ unsatisfied by }X)  \\
\le \P( S \text{ is $(T,\xi)$-bad}) 
+ 
\P(\text{all clauses in }T_{\good}\text{ unsatisfied by }X \mid S \text{ is not $(T,\xi)$-bad})  
\end{multline}
For $S$ that is not $(T,\xi)$-bad, one has
\begin{align*}
|\vbl(T_{\good}) \cap (\mcM \setminus S)| 
&\ge |\vbl(T_{\good}) \cap \mcM| - |\vbl(T_{\good}) \cap S| \\
&\ge \kalpha |T_{\good}| - \xi \theta k |T| \\
&\ge (1 - 6\zeta) \kalpha |T| - \xi \theta k |T|,  
\end{align*}
where the last inequality is due to $|T_{\good}| \ge (1 - 6\zeta)|T|$ from \cref{c:dtreeexists-conn}.
Hence, we deduce from the local uniformity of \cref{l:atklu} that the second term of \cref{eq:first-second} is upper bounded by
\begin{align*}
\left( \frac{1}{2} \exp\left(\frac{1}{ k}\right)\right)^{(1 -6\zeta) \kalpha |T| - \xi \theta k |T| }
&= \exp\left( - \left( \ln 2 - \frac{1}{k} \right)\left((1 -6\zeta)\kalpha - \xi\theta k \right) |T| \right) \\
&\le \exp\left( - \left( (\ln 2)(1 -6\zeta) \kalpha - (\ln 2) \xi\theta k - 1 \right) |T| \right),
\end{align*}
where we use $(1 -6\zeta)\kalpha - \xi\theta k \le (1 -6\zeta)\kalpha \le k$.

Next we bound $\P( S \text{ is $(T,\xi)$-bad})$, the first term in \cref{eq:first-second}. 
Observe that $\EE[|\vbl(T_{\good}) \cap S|] \in ((1 -6\zeta)\theta \kalpha |T|, \theta (k-\kbeta) |T|)$. 
Since $S$ is a subset of fixed size, the events $v \in S$ are negatively correlated for distinct $v \neq w \in \mcM$.
Therefore, we can estimate with a Chernoff bound that 
for sufficiently large $\xi \ge 1$ and $\eps \in (0, 1),$
\begin{align*}
\P( S \text{ is $(T,\xi)$-bad} ) &= \P( |\vbl(T_{\good}) \cap S| \ge  \xi \theta k |T|) \\
&\le \P( |\vbl(T_{\good}) \cap S| \ge  \xi \theta (k - \kbeta) |T|) \\
&\le \exp \left(- \frac{(\xi - 1)^2}{\xi + 1} \cdot (1 -6\zeta) \theta \kalpha |T| \right) \\
&\le \exp \left(- (1-\eps) (1 -6\zeta) \xi \theta \kalpha |T| \right),
\end{align*}
where the last inequality follows from $(\xi - 1)^2  \ge (1-\eps)\xi(\xi + 1)$ when $\xi \ge \xi(\eps)$. 

Recall that with high probability over $\Phi$ by \cref{l:numts}, the total number of choices of $T$ is crudely upper bounded by $\alpha n (18 k^2 \alpha)^{2|T|}$ since the number of good clauses is at most $\alpha n$ and $T \in \mcD^{(2)}(\Phi)$.
Letting $\eps = 0.01$ and $\xi = \frac{1}{\theta} \cdot \frac{2 \gamma (\ln 2) (1 - 6\zeta)(1 - \zeta)}{\ln 2 + 2 \gamma(1 - \eps)(1 - 6\zeta)(1-\zeta) }$
we see that for sufficiently small $\theta$ (so that $\xi \ge 300$), we have that
\begin{align*}
& \P(\mcB_t) \\
\le{}& \alpha n (18 k^2 \alpha)^{2|T|} \left[ \exp \left(- (1 - \eps)(1 - 6\zeta) \xi \theta \kalpha |T| \right) + \exp\left( - \left( \ln 2 \cdot (1 - 6\zeta)\kalpha - (\ln 2) \xi\theta k - 1 \right) |T| \right) \right] \\
\le{}& 3 \alpha n (18 k^2 \alpha)^{2|T|} \exp \left(- (1-\eps)(1 - 6\zeta) \xi \theta \kalpha |T| \right) \\
\le{}& \alpha n \cdot 2^{-|T|} 
= \frac{1}{n^{\Omega(\Delta k^2)}},
\end{align*}
where we also use that $k$ is sufficiently large and $\alpha \le  2^{k/25} \ll \frac{1}{k^2} 2^{ \frac{2\gamma^2 (1-\eps)(1 - 6\zeta)^2(1-\zeta)^2}{\ln 2 + 2\gamma(1-\eps)(1-6\zeta)(1-\zeta)} k}$.

Similarly, let $\mcB$ be the bad event that when we compute $X$ by extending $X_{T_{\max}}$ (an assignment on $\mcM$) to an assignment on all of $V$, some component of $G_{\Phi^{X_{T_{\max}}}}$ has size $\ge R \Delta^2 k^3 \log n$. 
By an identical argument to above, we have $\P(\mcB) = n^{-\Omega(\Delta k^2)}$. 
Since $T_{\max} = n^{O(\Delta k^2)}$, for sufficiently large constant $R$, we see that by taking a union bound, the probability of any bad event occurring is at most $O(1/n^2)$.
\end{proof}

\begin{rmk} \label{r:treeexcess}
Naively, searching over the connected components of size $O(\Delta^2 k^3 \log n)$ to sample the marginal probabilities takes $n^{O(\Delta^2 k^3)}$ time. However, we can observe that with high probability, logarithmically sized components have constant tree excess, which allows us to sample from marginal distributions in time $O(2^{2k} \poly(n))$. Unfortunately, the above approach is only able to show mixing in $O(n^{\Delta k})$ iterations, given the black box application of spectral independence. We suspect this algorithm mixes in $O_k(\poly(n))$ iterations and can be implemented in $O_k(\poly(n))$ time.
\end{rmk}

\subsection{Looseness for random \texorpdfstring{$k$}{k}-CNFs}
We next show $O(\log n)$-looseness for all variables with high probability over $(\Phi, \sigma)$ for random $k$-CNF instances $\Phi \sim \Phi_k(m, n)$ and uniformly random satisfying assignment $\sigma \in \Omega$. Consequently, in an \textit{algorithmic} regime where $d \ll 2^{ck}$ for some $c < 1$, we resolve a conjecture of~\cite{ACH08}. Our work further implies a reduction from looseness results of bounded degree CNFs to associated looseness for slightly sparser random $k$-CNFs.
Here, we employ the same notational conventions and parameter regime as in~\cref{s:randcon} with the slightly stronger assumption that $\alpha \le 2^{(1-9\zeta) \gamma k}$.

\begin{defn}
Let $\Phi \sim \Phi_k(m, n)$ be a random $k$-CNF. Variable $v \in V$ is \textit{flippable} if there exists a pair of satisfying assignments to $\Phi$, in one of which $X(v) = 0$ and in the other $X(v) = 1$.

For $\Phi \sim \Phi_k(m, n)$ as above and $\sigma \rand \Omega$ a uniformly random solution, we say that 
$\Phi_k(m, n)$ is \textit{$f(n)$-loose} if with high probability over $(\Phi, \sigma)$, in $\Phi$, all variables $v \in V$ are $f(n)$-loose with respect to $\sigma$.
\end{defn}

\begin{lemma}\label{l:mostflippable}
For $\alpha < 2^{k-1}$, with high probability over $\Phi \sim \Phi_k(m, n)$ all variables in $\Phi$ are flippable.
\end{lemma}
\begin{proof}
Observe that we can view $\Phi$ as an instance of NAE-SAT, in the sense that every clause $C$ forbids both $\varphi_C$, the original assignment that $C$ forbids, 
and $\overline{\varphi_C}$, the opposite of $\varphi_C$. 
By Theorem~2 in \cite{AM02}, with high probability $\Phi$ is NAE-satisfiable. 
Consequently, we can find some assignment $\sigma$ that NAE-satisfies $\Phi$ with high probability, and then the opposite assignment $\overline{\sigma} = \ind - \sigma$ also NAE-satisfies $\Phi$ by the symmetry of NAE-SAT solutions. 
In particular, both $\sigma$ and $\overline{\sigma}$ are solutions to the original SAT formula $\Phi$. 
Observe that for every variable $v \in V$ we have $X(v) = 1$ and $X(v) = 0$ in exactly one of $\sigma, \overline{\sigma}$ and thus every variable in $\Phi$ is flippable with high probability.
\end{proof}

\begin{lem}\label{lem:XY}
For any bad variable $v \in \mcV_{\bad}$ and any partial assignment $X \in \{0, 1\}^{\mcM}$, we have
\[
\mu_v(0|X) >0 
\quad\text{and}\quad
\mu_v(1|X) > 0.
\]
\end{lem}

\begin{proof}
To prove $\mu_v(0|X) > 0$ it suffices to show that
\begin{equation}\label{eq:XY}
\mu(\sigma(v) = 0, \sigma(\mcM) = X) > 0.
\end{equation}
By \cref{l:mostflippable} there exists a satisfying assignment $\sigma$ with $\sigma(v) = 0$. 
Let $Y = \sigma(\mcV_{\bad})$ be the assignment on bad variables and so in particular $\mu_{\mcV_{\bad}}(Y) > 0$.
Then $\Phi^Y$ is a $(k,\zeta,\Delta)$-CNF and by \cref{l:vtxapproxuniform} we know that for any $X \in \{0, 1\}^{\mcM}$ it holds that
\[
\mu_{\mcM}(X|Y) > 0.
\]
This implies that 
\[
\mu(\sigma(\mcM) = X, \sigma(\mcV_{\bad}) = Y) > 0,
\]
and in particular \cref{eq:XY} holds.
\end{proof}

\begin{proof}[Proof of \cref{t:randloose}]
Consider a random $k$-CNF $\Phi \sim \Phi_k(m, n)$ and a uniformly random satisfying assignment $\sigma$. Let $\Delta = k^4 \alpha$ as in~\cref{s:randcon}.
Fix a marking $\mcM \subset \mcV_{\good} \subset V$ as in~\cref{l:randmark}. 

Let $Y$ be an arbitrary assignment on all bad variables that is extendable to a full satisfying assignment in $\Omega$. 
First, notice that the uniform distribution $\mu(\cdot |Y)$ over all satisfying assignments conditioned on $Y$ is $k$-locally uniform for the marginal on $\mcM$ by~\cref{lem:local-uni-RCNF}. 
By~\cref{t:lllloose}, all good variables are simultaneously $O(\Delta k^2 \log n)$-loose whp for an arbitrary (feasible) $Y$ and a uniformly random solution to $\Phi^Y$. 
This immediately implies that all good variables are $O(\Delta k^2 \log n)$-loose whp in a uniformly random solution to $\Phi$. 
(Note that the looseness here is in a stronger form in the sense that to flip a good variable, one only needs to flip $O(\Delta k^2 \log n)$ good variables to maintain a satisfying assignment but no bad variables need to be changed.)

Next we consider the case where we wish to flip some bad variable $v \in \mcV_{\bad}$. 
Let $X = \sigma(\mcM)$ where $\sigma$ is a uniformly random satisfying assignment to $\Phi$.
We will flip $v$ by updating the connected component containing $v$ in $H_{\Phi^X}$. 
By~\cref{lem:XY}, for an arbitrary pinning of $\mcM$, there exists a satisfying assignment where $v$ is $0$ and also a satisfying assignment where $v$ is $1$.
In other words, $v$ is flippable in $\Phi^X$ and to flip the value of $v$ one only needs to update the connected component containing $v$ in $H_{\Phi^X}$ since different components are independent of each other. 
Hence, it suffices to upper bound the size of the connected component containing $v$ in $H_{\Phi^X}.$
Specifically, we show that the component size is $O(\Delta k^2 \log n)$ with probability $1-O(1/n^2)$. The desired looseness then immediately follows from the union bound. 

Let $\mcB$ be the bad event that $H_{\Phi^{X}}$ has some component of size at least $ R \Delta k \log n$ for sufficiently large constant $R$.
We shall use the same proof strategy as for \cref{l:appxblock} to show that $\Pr(\mcB) \le 1/n^2$ and then the theorem follows from the union bound. 
If $\mcB$ occurs, then there exists some connected component $\mcE^{X}$ (where we do not remove marked variables fixed by $X$ from these clauses) with $|\mcE^{X} | \ge R \Delta k \log n$, 
and we can find  $T \subset \mcE^{X}$ satisfying the conditions of~\cref{c:dtreeexists-conn} with $|T| = \ceil{R \log n}$. 
Let $T_{\good} = T \cap \mcC_{\good}$ be the set of good clauses in $T$. 
We obtain that
\[
\P(\mcB) 
\le \sum_{T\text{ satisfying~\cref{c:dtreeexists-conn}}} \P(\text{all clauses in }T_{\good}\text{ unsatisfied by }X).
\] 
Fix a $T \in \mcD^{(2)}(\Phi)$ satisfying \cref{c:dtreeexists-conn}, 
and observe that $|\vbl(T_{\good}) \cap \mcM| \ge \kalpha |T_{\good}| \ge (1 - 6\zeta) \kalpha |T|$ 
where the last inequality follows from \cref{c:dtreeexists-conn}.
Hence, we deduce from the local uniformity of $X$ that 
\begin{align*}
\P(\text{all clauses in }T_{\good}\text{ unsatisfied by }X)
&\le \left( \frac{1}{2} \exp\left(\frac{1}{ k}\right)\right)^{(1 - 6\zeta) \kalpha |T|}.
\end{align*}
In conjunction with~\cref{l:numts}, we get with high probability over $\Phi$ that 
\[
\P(\mcB) \le \alpha n (18 k^2 \alpha)^{2|T|} \left( \frac{1}{2} \exp\left(\frac{1}{ k}\right)\right)^{(1 - 6\zeta) \kalpha |T|} \le \alpha n \left(\frac{18^2 e k^4 \alpha^2}{2^{(1 - 6\zeta) \kalpha }} \right)^{|T|} \le \frac{1}{n^{R/2}},
\]
where the last inequality follows from the density bound $\alpha \le 2^{(1 - 9\zeta) \gamma k} \ll \frac{1}{k^2} 2^{\frac12(1 - 6\zeta) \kalpha}$ where the final inequality holds for $k$ sufficiently large.
The theorem then follows.
\end{proof}

\begin{rem} 
We are able to show a weaker looseness result in a slightly denser regime than~\cref{t:randloose}. Specifically, our analysis of looseness in bounded degree CNFs and bad components of random $k$-CNFs implies that for some $\gamma \ge 0.1742$ and any $\zeta > 0$, there exists constant $c > 0$ such that if $\alpha \le \frac{c}{k^4} 2^{(1-\zeta)\gamma k}$, $\Phi_k(m, n)$ is $O(\log^3 n)$.
\end{rem}

We conjecture that $\Phi_k(m, n)$ is actually $O_k(\log n)$-loose in a larger regime, per~\cref{c:loose}.

\bibliographystyle{alpha}
\bibliography{ksat.bib}

\appendix

\section{Bad variables and components in random \texorpdfstring{$k$}{k}-CNFs}\label{a:badcomps}

Here, we prove~\cref{l:randbadcomp}, generalizing Lemma 48 of~\cite{GAL19}. In order to do this, we will need better understand the graph structure of random $k$-CNFs. 
Throughout, we employ the notation of~\cref{n:rand} and let $\zeta \in (0, 1/2)$ be a fixed small constant. Much of this generalizes work in  \S8 of~\cite{GAL19}.

Recall that a bad component is a connected component of $H_{\Phi, \bad}$, the graph where variables are connected if they appear together in a bad clause. Thus, $V(H_{\Phi, \bad}) = \mcV_{\bad}.$ 

\begin{defn}
Recall that $\HD(\Phi)$ are the high degree variables of $\Phi$. More generally, let $\HD(S) = \HD(\Phi) \cap S$ be the high degree variables in $S$.

Let $\BC(S)$ be the set of variables defined via the following iterative process:
\begin{itemize}
    \item Initialize: $\BC(S) = S$
    \item While there is a clause $C$ with $|\vbl(C) \cap \BC(S)| \ge \zeta k$ and $\vbl(C) \setminus \BC(S) \neq \emptyset$, let $\BC(S) \leftarrow \BC(S) \cup \vbl(C)$
\end{itemize}
\end{defn}

By construction $\mcV_{\bad} = \BC(\HD(\Phi))$ and for every bad component $S$, $S = \BC(\HD(S))$.

We first observe that $\Phi$ cannot have too many high degree variables, noting that $\alpha = m/n \le  \frac{1}{k^3} \Delta$. We first make the following standard observation.

\begin{lemma}\label{l:nhigh}
With high probability over $\Phi$, $|\HD(\Phi)| \le n e^{-\frac12 \Delta}.$
\end{lemma}
\begin{proof}
The degree distribution of the variables of $\Phi$ is given by a the distribution of randomly allocating $km$ balls (clause variables) into $n$ bins (the variables). Consequently, if $D_1, \ldots, D_n \sim \Poi(k\alpha)$ are independent Poisson random variables with rate $k\alpha$, then the distribution of the variable degrees in $\Phi$ is the same as the distribution of $\{D_1, \ldots, D_n\}$ conditioned on the sum $D_1 + \cdots + D_n = km$. Further, 
$\P(D_1 + \cdots + D_n = km) = O(1/\sqrt{n}).$
Thus, if $B = \{i \in [n] : D_i \ge \Delta\}$ we can estimate the Poisson tails to find that since $\alpha \le  \frac{\Delta}{k^3}$
$$\EE[|B|] = n \P(\Poi(k\alpha) \ge \Delta) \le n \exp\left(-\frac{(\Delta - k\alpha)^2}{\Delta} \right) \le n e^{-(1 - \frac{2}{k^2})\Delta}.$$
Then, by a Chernoff bound, we see that $\P(|B| \ge 2 \EE[|B|]) = \exp(-\Omega(n))$. Thus,
$$\P_{\Phi}(|\HD(\Phi)| \ge n e^{-\frac12 \Delta}) \le \P(|B| \ge 2 \EE[|B|] \mid D_1 + \cdots + D_n = km) = \exp(-\Omega(n)).$$
\end{proof}

We use the following pair of lemmas of~\cite{GAL19}, showing expansion bounds on small subset sof variables in a random formula $\Phi$
\begin{lemma}[Lemma 36~\cite{GAL19}]\label{l:smallexpansion}
Let $2 \le b \le k$ be an integer with $t = \frac{2}{b-1}$. With high probability over $\Phi$, for every set of variables $Y$ of size $2 \le |Y| \le \frac{n}{2^k}$, the number of clauses of $\Phi$ that have at least $b$ variables from $Y$ is at most $t |Y|$.
\end{lemma}

\begin{lemma}[Lemma 41~\cite{GAL19}]\label{l:technical}
Let $\delta_0 > 0$ and $\theta_0 \ge \min(k^2 \alpha, 2)$ be constants such that $\delta_0 \theta_0 \log(\theta_0/k^2 \alpha) > \log \alpha + 3\log k$. Then with high probability over $\Phi$, there do not exist sets $Y, Z$ of clauses and a set $U$ of variables such that 
\begin{itemize}
    \item $|Y| \ge \log n, |U| \ge \delta_0 |Y|, |Z| \ge \theta_0 |U|$, $Y \cap Z = \emptyset$
    \item $G_{\Phi}[Y]$ is connected, $U \subset \vbl(Y)$ and every clause in $Z$ contains at least one variable from $U$.
\end{itemize}
\end{lemma}

\begin{lemma}\label{l:boundhighdeg}
Fix constant $\xi > 5$. With high probability over $\Phi$, every connected set $U$ of variables with size at least $\xi k \log n$ has $|\HD(U)| \le \frac{|U|}{\xi}$
\end{lemma}
\begin{proof}
(This follows similarly to Lemma 42 in~\cite{GAL19}).
Let $\delta_0 = \frac{1}{\xi}$ and $\theta_0 = \Delta - 2(k+1)$. Observe that for sufficiently large $k$, $$\delta_0 \theta_0 \log \frac{\theta_0}{k^2\alpha} \ge \frac{1}{\xi} k(\Delta - 2(k+1)) \ge k + 3 \log k \ge \log \alpha + 3 \log k.$$
Consequently, $\Phi$ satisfies the conditions of~\cref{l:technical}.

Suppose to the contrary, we could find a bad connected set of variables $B$ with $|B| \ge \xi k \log n$ such that $|\HD(B)| > \frac{1}{\xi}|B|$. 

Consider the bipartite factor graph $F_{\Phi}$ with a vertex for each variable and clause and an edge for each variable-clause incidence.

Since $H_{\Phi}(B)$ is connected, we can find a tree $T' \subset F_{\Phi}$ with at most $2|B|$ vertices that includes all variables in $\HD(B)$ and at most $|B|$ clauses. Then, for any clause $C \in T'$ that contains some variable in $\HD(B)$ but does not have any neighbors in $T'$ in $\HD(B)$, we do the following replacement. Let $b \in C \cap \HD(B)$, and let $v \in V \setminus \HD(B)$ be a neighbor in $T'$ of $C$ on the path $C \rightarrow v \rightarrow \cdots \rightarrow b$. We can then remove edge $(C, v)$ from $T'$ and replace it with $(C, b)$. Iterating, we may assume that for any clause $C \in T'$ with some variable in $\HD(B)$, at least one of $C$'s neighbors is in $T' \cap \HD(B).$ Further, we can prune any leaves not in $\HD(B)$, so that the leaves of our resulting tree $T \subset T' \subset F_{\Phi}$ are a subset of $\HD(B)$ and $T$ has at most $2|B|$ vertices and includes all variables in $\HD(B)$ and at most $|B|$ clauses. Let $T_L$ be the set of leaves of $T$.

Let $\mcC(T)$ be the set of clauses of $T$. We note that
$$|\HD(B)| \le k|\mcC(T)| \le k|B|$$
By assumption on $B$, $|\HD(B)| > \frac{1}{\xi}|B| \ge k \log n,$ so $|\mcC(T)| \ge \log n.$ Each variable in $\HD(B)$ is contained in at least $\Delta$ clauses. Further, $|\HD(B)| \le |\HD(\Phi)| \le \frac{n}{2^k}$, and thus by applying~\cref{l:smallexpansion} with $b = t = 2$, we see that the number of clauses that contain at least $2$ variables from $B$ is at most $2|\HD(B)|.$
Consequently, the number of clauses that contain at least one variable from $\HD(U)$ is at least $\Delta |\HD(U)| - 2 |\HD(U)|k.$

Now suppose that $\eta$ clauses in $\mcC(T)$ contain at least one variable in $\HD(B)$. By construction,
$$\eta \le \sum_{b \in \HD(B)} \deg_T(b)$$
If $D = \sum_{b \in \HD(B) \setminus T_L} \deg_T(b)$, since $T$ is a tree containing all of $\HD(B)$, we see that 
$$|T_L| = 2 + \sum_{v \in T \setminus T_L} (\deg_T(v) - 2) \ge 2 + \sum_{v \in \HD(B) \setminus T_L} (\deg_T(v) - 2) = 2 + D - 2(|\HD(B)| - |T_L|).$$
Consequently, 
$$\eta \le D + |T_L| \le 2 |\HD(B)|,$$
i.e. at most $2|\HD(B)|$ clauses in $T$ contain a variable from $|\HD(U)|$. 

Therefore, there exists a set $Z$ of clauses of size at least $(\Delta - 2k - 2)\HD(B) = \theta_0 |\HD(B)|$ that contain a variable in $\HD(B)$ but where $Z \cap \mcC(T) = \emptyset.$

By construction $|\mcC(T)| \ge \log n, |\HD(B)| \ge \delta_0|\mcC(T)|, |Z| \ge \theta_0|\HD(B)|$. Since $\mcC(T)$ is connected and $\vbl(\mcC(T)) \supset \HD(B)$. Every clause in $Z$ contains at least one variable from $\HD(B)$. However, then $\Phi$ does not satisfy \cref{l:technical}. This proves the result.
\end{proof}

We leverage an expansion property of $\Phi$ to control the number of bad variables.

\begin{lemma}[Lemma 2.4~\cite{CF14}]\label{l:exp2}
There exists a constant $k_0 > 0$ such that for all $k \ge k_0$ the following holds. With probability $1 - o(1/n)$ over the choice of the random formula $\Phi$, for $\eps > 0$ and $\lambda > 4$ satisfying $\eps \le k^{-3}$ and $\eps^{\lambda} \le \frac{1}{e} (2e)^{-4k}$, $\Phi$ has the following property.

Let $Z \subset [m]$ is any set of size $|Z| \le \eps n$ and $i_1, \ldots, i_{\ell} \in [m] \setminus Z$ be a sequence of distinct indices. For $s \in [\ell]$, let $N_s := \vbl(Z) \cup \bigcup_{j = 1}^{s-1} \vbl(C_{i_j})$. If
$$|\vbl(C_{i_s}) \cap N_s| \ge \lambda, \quad \forall s \in [\ell],$$
then $\ell \le \eps n.$
\end{lemma}

We deduce the following corollary:
\begin{cor}\label{c:fewlargeoverlap}
With high probability over $\Phi$, if $Z \subset [m]$ is a set of size $\le 2 n e^{-\frac12 \Delta}$ and $i_1, \ldots, i_{\ell} \in [m] \setminus Z$ is a sequence  of distinct indices. For $s \in [\ell],$ let $N_s := \vbl(Z) \cup \bigcup_{j = 1}^{s-1} \vbl(C_{i_j})$. If
$$|\vbl(C_{i_s}) \cap N_s| \ge \zeta k, \quad \forall s \in [\ell],$$
then $\ell \le |Z|$
\end{cor}
\begin{proof}
For an integer satisfying $1 \le z \le  2 n e^{-\frac12 \Delta}$, let $\mcE_z$ be the event that there exists a set $Z$ with $|Z| = z$ that does not satisfy the desired property. Then, we apply \cref{l:exp2} with $\eps = z/n$ and $\lambda = \zeta k$. Then, $\eps = \frac{z}{n} = 2e^{-\frac12 \Delta} \le k^{-3}$ and 
$$\eps^{\lambda} = \left(2e^{-\frac12 \Delta}\right)^{\zeta k} = 2^{\zeta k} e^{-\frac12 \Delta \zeta k} \le e^{-8k-2} \le \frac{1}{e} (2e)^{-4k}.$$
Consequently, $\P_{\Phi}(\mcE_z) = o(1/n)$ and thus via a union bound over $z$, we get the desired result.
\end{proof}

\begin{lemma}\label{l:fewbad}
With high probability over $\Phi$, $|\mcV_{\bad}| \le 4k n e^{-\frac12 \Delta}$
\end{lemma}
\begin{proof}
With high probability, $\Phi$ satisfies the properties in~\cref{l:nhigh,l:smallexpansion,c:fewlargeoverlap}. Thus, $|\HD(\Phi)| \le n e^{-\frac12 \Delta}$ and thus by applying~\cref{l:smallexpansion} with $b = t = 2$, we see that the number of clauses with at least $2$ high degree variables is at most $2|\HD(\Phi)| \le 2n e^{-\frac12\Delta}$ and thus $|\mcC_0| \le  2n e^{-\frac12\Delta}$ (the first step in constructing high-degree variables). Then, starting with $\mcC_0$, by~\cref{c:fewlargeoverlap}, $|\mcC_{\bad}| \le 4n e^{-\frac12 \Delta}$ and thus $|\mcV_{\bad}| \le 4kn e^{-\frac12 \Delta}$
\end{proof}

\begin{lemma}\label{l:badcompsmall}
With high probability over $\Phi$, for any bad component $S$, $|S| \le \frac{6}{\zeta} |\HD(S)|.$
\end{lemma}
\begin{proof}
with high probability, $\Phi$ satisfies the properties we need for~\cref{l:nhigh,l:smallexpansion,c:fewlargeoverlap}. Suppose these properties are satisfied and let $S$ be a bad component. If $S$ consists of a single isolated variable, then $HD(S) = S$ trivially.

Else, $S$ is a connected component of variables in $H_{\Phi, \bad}$ and thus $S$ has at least $\zeta k$ high degree variables. Further, $|\HD(S)| \le |\HD(\Phi)| \le 2n e^{-\frac12 \Delta}.$ Applying~\cref{l:smallexpansion} with $b = \zeta k$, we see that the number of clauses with at least $\zeta k$ variables from $\HD(S)$ is at most $\frac{3}{\zeta k}|\HD(S)|$.

Now let's try to recursively extend $\HD(S)$ to the associated set of bad variables. Let $Z$ be the clauses with at least $\zeta k$ variables from $\HD(S)$ (so $|Z| \le \frac{3}{\zeta k} |\HD(S)| \le \frac{6n e^{-\frac12 \Delta}}{\zeta k}$.
Then, by~\cref{c:fewlargeoverlap}, the number of clauses $C$ with $\vbl(C) \subset \BC(\HD(S)) = S$ is at most $2|Z| \le \frac{12 n e^{-\frac12 \Delta}}{\zeta k}$. Since each variable in $S$ is contained in some bad clause, we see that 
$$|S| \le \left|\bigcup_{C \in \mcC_{\bad}} \vbl(C) \right| \le  \frac{6}{\zeta k} |\HD(S)| \cdot k \le \frac{6}{\zeta} |\HD(S)|$$
\end{proof}

\begin{proof}[Proof of \cref{l:randbadcomp}]
Let $\xi \ge \frac{7}{\zeta}$.
Suppose there is a bad component of size $|S| > \xi k\log n$, Since $S$ is a connected component in $H_{\Phi, \bad},$ $S$ is also connected in $H_{\Phi}$. We know by~\cref{l:boundhighdeg}, $|\HD(S)| \le \frac{1}{\xi}|S|$. However, by \cref{l:badcompsmall}, we see that $|\HD(S)| \ge \frac{\zeta}{6}|S| > \frac{1}{\xi}|S|,$ a contradiction.
\end{proof}

\section{Missing proofs in~\texorpdfstring{\cref{sec:random-CNF}}{Section 3}}\label{a:missingsec3}

To show~\cref{l:fewbadclauses} we begin with a helpful intermediate lemma, building on the results of~\cref{a:badcomps}.
\begin{lemma}\label{l:fewbadvars}
With high probability over $\Phi$, for every connected set $U$ in $H_{\Phi}$ with $|U| \ge \frac{8}{\zeta^3} k \log n$, we have that $|U \cap \mcV_{\bad}| \le \zeta^2 |U|$.
\end{lemma}
\begin{proof}
With high probability, $\Phi$ satisfies the conditions of~\cref{l:badcompsmall,l:boundhighdeg}.
Suppose to the contrary that we could find some connected $U$ of size $|U| \ge \frac{8}{\zeta^3}  k \log n$ with $|U \cap \mcV_{\bad}| > \zeta^2 |U|$.
Let us suppose that there are $t$ bad components that intersect $U$, denoted by $S_1, S_2, \ldots, S_t$. Let $U' = U \cup S_1 \cup \cdots \cup S_t$ and let $f = \frac{|U' \backslash U|}{|U|}$. Since $U'$ is a connected set of variables and all variables in $U' \backslash U$ are bad, we observe that $|U' \cap \mcV_{\bad}| \ge (\zeta^2 + f)|U|$. Since every bad variable in $U$ is contained in exactly one $S_i$, we apply~\cref{l:badcompsmall} to see that
\footnotesize
$$|\HD(U')| = \sum_{i = 1}^t |\HD(S_i)| \ge \frac{\zeta}{6} \sum_{i = 1}^t |S_i| = \frac{\zeta}{6} |U' \cap \mcV_{\bad}| \ge \frac{\zeta}{6} \left(\zeta^2 + f\right)|U| = \frac{\zeta^3}{6} \left(1 + \frac{f}{\zeta^2}\right)|U| \ge \frac{\zeta^3}{6} (1 + f)|U| = \frac{\zeta^3}{6} |U'|.$$
\normalsize
Note however, that $|U'| \ge |U| \ge \frac{8}{\zeta^3} k \log n$, and thus by~\cref{l:boundhighdeg} with $\xi = \frac{8}{\zeta^3}$, we arrive at the contradiction that $|\HD(U')| \le \frac{\zeta^3}{8}|U'|$. This implies the desired result.
\end{proof}

\begin{proof}[Proof of~\cref{l:fewbadclauses}]
With high probability, we have that $\Phi$ satisfies the conditions of~\cref{l:fewbadvars,l:fewbad,l:smallexpansion}. Take $Y$ as above, and let $U = \vbl(Y)$, so $U$ is a connected set of size $|U| \ge \frac{8}{\zeta^3} k \log n$. By~\cref{l:fewbadvars}, we then have $|U \cap \mcV_{\bad}| \le \zeta^2 |U|.$

By~\cref{l:fewbad}, $|U \cap \mcV_{\bad}| \le |\mcV_{\bad}| \le 4 k n e^{-\frac12 \Delta}$ and thus we can apply~\cref{l:smallexpansion} with $b = \zeta k$ in conjunction with the above bound to obtain the desired upper bound on the number of clauses of $Y$ with at least $\zeta k$ variables from $\mcV_{\bad}$:
$$|Y \cap \mcC_{\bad}| \le \frac{3}{\zeta k}|U \cap \mcV_{\bad}| \le 3 \zeta \frac{|U|}{k} \le 3 \zeta |Y|.$$
This establishes the lemma.
\end{proof}

\begin{proof}[Proof of~\cref{c:dtreeexists-conn}]
We apply~\cref{l:greenblue} to $G_\Phi[\mcE_0]$, coloring good clauses (vertices) green, bad clauses blue, and coloring edges between (good) clauses that share a good variable green and edges between clauses that share a bad variable blue. Note that every bad clause has only bad variables and every good clause has at most $\zeta k$ bad variables. In this coloring, $V_{\mathrm{g}} = \mcC_{\good} \cap \mcE_0$ and $V_{\mathrm{b}} = \mcC_{\bad} \cap \mcE_0$.
Thus, we can apply~\cref{l:greenblue} with $D = (\Delta - 1)k$, to find a subset of clauses $T \subset \mcE_0 \subset  V(G_{\Phi}) = \mcC(\Phi)$ such that 
\begin{enumerate}
    \item $\mcC_{\bad} \cap \mcE_0 \subset T$;
    \item $T \cap \mcC_{\good}$ is an independent set in $G_{\Phi,\good}$ with $$|T \cap \mcC_{\good}| \ge \frac{|\mcC_{\good} \cap \mcE_0|}{(\Delta - 1) k + 1} \ge \frac{|\mcE_0 \cap \mcC_{\good}|}{\Delta k};$$
    \item $G_{\Phi}^{\le 2}[T]$ is connected.
\end{enumerate}
We will use these properties to verify that $T$ satisfies the conditions of the lemma.
\begin{enumerate}[(a)]
    \item The above conditions on $T$ from~\cref{l:greenblue} immediately imply that $T \in \mcD^{(2)}(\Phi)$.
    \item If $|\mcE_0| \ge R\Delta k \log n$, then since
$$|T| = |\mcE_0 \cap \mcC_{\bad}| + |I| \ge |\mcE_0 \cap \mcC_{\bad}| + \frac{|\mcE_0 \cap \mcC_{\good}|}{\Delta k} \ge \frac{|\mcE_0|}{\Delta k} \ge R \log n,$$
we can take a spanning tree of $G_{\Phi}^{\le 2}[T]$ and remove leaf vertices to assume without loss of generality that $|T| = \ell.$
    \item  Finally, we show that $|T \cap \mcC_{\good}| \ge (1 - 6\zeta)|T|$. 
 Since $T \in \mcD^{(2)}(\Phi)$, \cref{l:conind} implies that we can find a set $T \subset T' \subset V(G_{\Phi})$ with $|T'| \le 2|T|$ such that $G_{\Phi}[T']$ is connected. By \cref{l:expstrong},
 $$|\vbl(T')| \ge 0.9k|T'| \ge 0.9k|T| \ge 0.9R k \log n.$$
 By letting $R \ge \frac{80}{9 \zeta^3}$ be a sufficiently large constant, we can apply \cref{l:fewbadclauses} and observe that
 \begin{equation*}
     |T \cap \mcC_{\bad}| \le |T' \cap \mcC_{\bad}| \le 3\zeta |T'| \le 6 \zeta |T|.
 \end{equation*}
 Hence, we obtain $|T \cap \mcC_{\good}| \ge (1 - 6\zeta)|T|$. \qedhere
\end{enumerate}
\end{proof}

\end{document}